  \documentclass[iop]{emulateapj}   % use this for ApJ
\usepackage{graphics}
 \usepackage{epsfig}
\usepackage{psfig}

\def\gtrsim{\mathrel{\hbox{\rlap{\hbox{\lower4pt\hbox{$\sim$}}}\hbox{$>$}}}}
\def\ltsim{\mathrel{\hbox{\rlap{\hbox{\lower4pt\hbox{$\sim$}}}\hbox{$<$}}}}

\slugcomment{Version of \today}
\shortauthors{SMITH, BIANCHI, \& SHIAO}
\shorttitle{GALEX-SDSS COLOR DIAGRAMS OF STARS}

\begin{document}

\newcommand{\teff}{T$_{\rm eff}$}
\newcommand{\logg}{log\,g}

\newcommand{\fuse}{{\it FUSE}}

\title{
Interesting Features in the Combined GALEX and Sloan Color Diagrams of 
Solar-like Galactic Populations
}

\author{MYRON A. SMITH}
\affil{Space Telescope Science Institute,\\
        3700 San Martin Dr., \\
        Baltimore, MD 21218;           \\
        myronmeister@gmail.com}

\author{LUCIANA BIANCHI}
\affil{Johns Hopkins University, \\
3700 San Martin Dr., \\
Baltimore, MD, 21218; \\
bianchi@pha.jhu.edu }

\and

\author {BERNARD SHIAO}
\affil{Space Telescope Science Institute,\\
        3700 San Martin Dr., \\
        Baltimore, MD 21218;           \\
        shiao@stsci.edu}

\begin{abstract}

We report on intriguing photometric properties of Galactic stars observed
in the GALEX satellite's far-UV (FUV) and near-UV (NUV) bandbasses as well as 
from the ground-based SDSS survey and the Kepler Input Catalog. The first 
property is that the (FUV-NUV) color distribution of stars in the 
Kepler field consists of two well separated peaks. A second and the more
perplexing property is that for stars with spectral types G or later the 
mean (FUV-NUV) color becomes much bluer, contrary to expectation. 
Investigating this tendency further, we found in two samples of mid-F through 
K type stars that 17-22\% of them exhibit FUV-excesses relative to their NUV 
fluxes and spectral types. A correction for FUV incompleteness of the FUV 
magnitude limited
star sample brings this ratio to 14-18\%. Nearly the same fractions 
are also discovered among members of the Kepler Eclipsing Binary Catalog and 
in the published list of Kepler Objects of Interest. These UV-excess (``UVe") 
colors are confirmed by the negative UV continuum slopes in GALEX spectra 
of members of the population.
The SDSS spectra of some UVe stars exhibit metallic line weakening, 
especially in the blue.  This suggests an enhanced contribution of UV
flux relative to photospheric flux of a solar-type single star.
We consider the possibility that the UV excesses originate from various
types of hot stars, including white dwarf DA and sdB stars, binaries, and 
strong chromosphere stars that are young or in active binaries. 
The space density of compact stars is too low to explain the observed 
frequency of the UVe stars. Our model atmosphere-derived 
simulations of colors for binaries with main sequence
pairs with a hot secondary demonstrate that the color loci conflict with the 
observed sequence. As a preferred explanation we are left with the 
active chromospheres explanation, whether in active close binaries or young 
single stars, as a still tentative explanation for the UVe population - despite
the expected paucity of young, chromospherically-active stars in the field.
% In addition, the latter two explanations present problems 
% with our current understanding of important phases of stellar evolution.
We also address a third perplexing color property, namely the presence of 
a prominent island of ``UV red" stars surrounded by ``UV blue" stars 
in the diagnostic $(NUV-g)$, $(g-i)$ color diagram. We find that the
subpopulation comprising this island are mainly horizontal branch stars. 
These objects do not exhibit UV excesses and therefore have UV colors
typical for their spectral types. This subpopulation appears ``red" in the 
UV only because their colors are not pulled to the blue by the inclusion
of UVe stars.
% in the total population.

\end{abstract}

\section{Introduction}

Although the Kepler program was conceived as a NASA space-borne mission 
dedicated to the discovery of Earth-sized exoplanets, its data archive is 
now being mined as an important resource for determining fundamental
properties of late-type stars, including their evolution and variability,
as well as a resource for the discovery of new stellar populations. 
The ensuing literature promises to mark the Kepler mission's 
% 105 square degree 
Field of View (FOV) as a region 
of dedicated study of Galactic disk stars for some time to come.

The study of Kepler light curves has required complementary ground-based
observations to determine basic stellar parameter. To take one 
%(mass, \teff, \logg, and radius) - a planetary host star's radius
example, a planetary host star's radius is used with the 
eclipse depth to determine the radius of a transiting planet.
% measurement of a planet's radius through the light curve
%transient depends upon the star's radius
%essential to determining the radii and densities of its planets.
The first large ground-based effort was the construction of 
the Kepler Input Catalog (KIC), which contained mean magnitudes obtained 
from the clones of the ${\it g, r, i, z}$ filters introduced by the Sloan
Digital Sky Survey (SDSS) project\footnote{The SDSS is managed by the 
Astrophysical Research
Consortium for the Participating Institutions. Funding for SDSS was
provided by the Alfred P. Sloan Foundation, Participating Institutions, the 
National Science Foundation, U.S. Department of Energy, and the National 
Aeronautics \& Space Administration. The 2.5m SDSS telescope is described
by Gunn et al. (2006). Technical summary of the SDSS project are given by
York et al. (2000) and Stoughton et al. (2002).  } 
before the launch of the Kepler spacecraft. 
Described by T. Brown et al. (2011), the KIC provided optical 
and near-IR magnitudes for nearly all stars down to magnitude 
17 in the region of the sky where Kepler's camera would be pointed. 
The ``Kepler magnitude" (K$_p$) used by the KIC is a weighted mean 
of the $g$ and $r$ magnitudes.
The catalog also included ${\it JHK}$ 
infrared observations from the {\it 2MASS} survey. 
Because foremost the SDSS was a survey of extragalactic objects, 
the sky coverage of the principal SDSS data releases avoided low
Galactic latitude sky zones and this avoidance included the Kepler FOV. 
The FOV is a 105 deg.$^{2}$ field and
%in one of these zones that is, the subregions included by pre-launch
%ground-based observations by the KIC program. most of which
% was accessible to the 42 CCDs of the Kepler imaging camera. 
%The Kepler FOV 
lies within  Galactic latitudes 5-22$^{\circ}$ 
toward the constellations Lyra and Cygnus. 
%along the Orion arm.
 
  Building on the success of the SDSS project, Groot et al. (2009) and Verbeek 
et al.  (2012) published the ``UVEX catalog," a SDSS filter-based photometric 
survey of objects with strong UV excess fluxes in a 1850 deg.$^{2}$ region of 
the Northern Galactic plane.  Steeghs et al. (2012) used SDSS filter copies
to observe 98\% of the area comprising the Kepler FOV, which as we have noted 
likewise substantially 
restricts studies to stars in the Galactic (thin) disk. This work is known as
the Kepler Isaac Newton Telescope Survey (``KIS," Steeghs et al. 2012), is
calibrated in the Vega magnitude system, and is distributed by MAST. 
%(http://archive.stsci.edu)  in two public releases. 
Our work includes the results of both releases, that is the union of both 
sky areas.  A spectroscopic follow up to the UVEX catalog by 
Verbeek et al. (2012b) confirmed the catalog authors' conclusions that 95\% 
of UVEX objects are hot objects such as DA and DB 
white dwarfs (WD), WD binaries, hot subdwarf (sdO, sdB) stars, and QSOs. 
Rounding out the other 5\% are upper main sequence stars and members
of the Blue and Extreme Horizontal Branches. 
%(BHB, eHB). 
%The authors found that 47\% of the UVEX population are DA WDs. 
%The WDs are mainly faint objects ($g$ $\gtrsim$17), whereas the 
%sdOB stars tend to be somewhat brighter.  
%%It is also important to point out for later that the hot subdwarf 
%%(sdB) sequence is shifted by $\approx$ -0.7 mags.in $(U-g)$ in the 
%%$g-r, U-g$ diagram at a given optical $(g-r)$ color.

  The addition of UV sky surveys from the GALEX satellite 
extended the wavelength coverage to simultaneous
observations in the near-UV (NUV, centered at 2271\,\AA) 
and far-UV (FUV, centered at 1528\,\AA) wavebands. The photometric depth of 
these surveys depended on the wavelength band and exposure time associated 
with the survey (e.g., Morrissey et al. 2007, Bianchi 2009, 2011).
Like the SDSS the GALEX surveys avoided the low Galactic
latitude sky regions owing to safety limits imposed by diffuse glow 
and bright stars (m$_{NUV} <$ 9.5). 
%However, building on GALEX's Guest Investigator and Nearby Galaxies
%Survey efforts, additional GALEX observations were made of the Kepler FOV.  
%Roughly half the FOV region had been observed in both NUV and FUV by the
%time the NASA phase of the GALEX mission terminated in early 2012. 
For completeness, we note that an electrical overcurrent in the FUV
camera caused its shutdown in May, 2009. NASA funding for the
GALEX mission was terminated in early 2012.
%, and the satellite ceased operations altogether in April, 2013. 
As for the Kepler spacecraft, the
failure of a reaction wheel in May 2013 brought to a close its search 
for exoplanets in the FOV. 

  Despite the extensive GALEX literature, rather few studies 
of stellar properties have been published to date that combine 
optical, IR, and GALEX photometry in the Kepler FOV. 
One reason for this still incomplete record is
%, ironically, the large
%quantity of KIC photometric optical/IR colors that characterize the properties 
%of planet-hosting cool stars. 
that whereas IR magnitudes form an important
complement, the UV magnitudes are faint and are generally less important 
to characterizing a cool star. Also, because of the diffuse UV glow in the
Galactic Plane, a homogeneous GALEX survey could not cover more than half 
the Kepler FOV. 
The optical and GALEX imaging surveys can accumulate strong UV fluxes
of several important stellar populations which are largely unrelated to
Kepler's core mission studies.  By making use of the (UV/optical) $U$ filter, 
KIS provides a bridge between the KIC optical and the GALEX NUV 
and FUV magnitudes and facilitates the discovery of Galactic stars with UV
excesses in this important sky field. 
%indeed whether they happen to be planet-hosting stars or not.

   Several studies that combine SDSS and GALEX data have contributed to our
understanding of stellar populations, particularly by deriving photometric 
%context of utilizing photometric colors to provide ersatz
spectral types for Galactic stars.  
Bianchi et al. (2005, 2007, 2011) and Bianchi (2011) have published 
initial results from these comprehensive photometric surveys. 
These authors computed colors of galaxies, QSOs, and well represented 
populations of Galactic stars (main sequence, giants/supergiants, white dwarfs)
through the SDSS and GALEX filters. This body of work demonstrated that 
unresolved galaxies and QSOs can be differentiated well by using GALEX 
UV colors combined with the SDSS ground-based data.
Comparison of the GALEX and SDSS surveys shows that in both the
Medium and All-Sky Imaging Surveys, the number of Galactic stars per
square degree falls off and equals the number (rapidly rising with magnitude) 
of GALEX-identified galaxies at K$_p$ = 17-18 magnitude. 

 While the addition of GALEX UV colors has facilitated searches 
for hot objects with UV excesses, new binary classes have been found that
include cool companions.
%SDSS survey.  This is true for cool as well as hot WDs (Bianchi et al. 2007, 
%2011a, 2011b, Groot et al. 2009, Girven et al. 2012a, Verbeek 2012a, 2012b).
Thus, GALEX and SDSS colors together have enabled the discovery 
of White Dwarf-Main Sequence systems (WDMS), i.e., binaries with 
WD primaries and late-type 
main sequence secondaries (see Rebassa-Mansergas et al. 2010, 2012,
Girven et al. 2012b, N\'emeth et al. 2012). 
%In addition, GALEX-SDSS photometry have made possible
%the identifications of faint Asymptotic Giant Branch 
%(Sahai et al. 2008) and main sequence Barium stars (Gray et al. 2011). 

  Herein we extend previous work by examining a newly discovered population 
of Galactic disk stars with GALEX far-UV excesses. Although these stars were 
first discovered in sky areas common to both GALEX and SDSS, our emphasis 
will be on the properties of similar objects that lie within the Kepler FOV
chiefly because of the promised rich harvest from ground-based follow-up 
campaigns on them. Thus, primarily we relied on the stars observed by the
GALEX GR6/GR7 surveys in the FOV. Secondarily, we relied on stars observed
by GALEX and SDSS DR7 imaging surveys, i.e. in sky regions not covered by 
the FOV and used by Smith \& Shiao (2011). 
Moreover, the SDSS filter colors used in each case are taken from different
magnitude systems. Because the samples are independent, we will treat
them separately. Both samples were created by adopting cross-correlation 
search radii of 5 arcsec and restricting matches that were closest neighbor
members of one survey to the other and vice versa.
The magnitude and error limits are discussed below.
% CHECK ON THIS goldstandard STATEMENT
Also, unless stated otherwise we generally constrained our study of the 
Kepler FOV to objects considered probable stars brighter than K$_p$ = 16.0 
and within 0.55\,$^{\circ}$ of the GALEX tile centers.\footnote{For the
SDSS/GALEX population the faint limit was K$_p$= 18.} This is also the 
practical limit for KIS recommended by Steeghs et al. (2012).

\section{Statement of the questions }

\begin{figure}[ht!]
\centerline{
\includegraphics[scale=.40,angle=90]{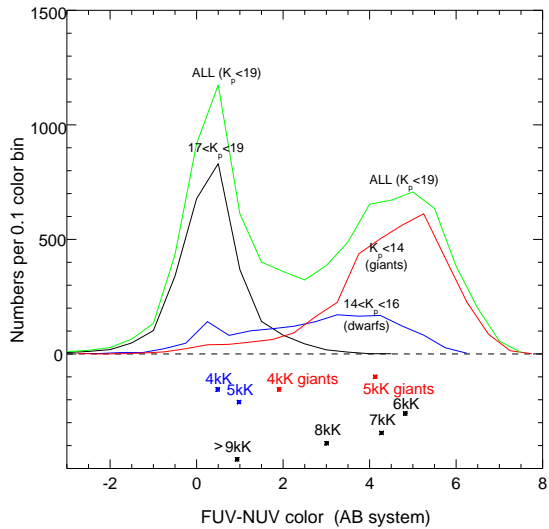}}
%alias histofncolddwfgnt.ps
\caption{GALEX (FUV-NUV) color distributions differentiated
by Kepler magnitude (K$_{p}$) and KIC \logg\, for dwarfs and giants.
These stars have been observed by GALEX in the Kepler Field of View. 
%The populations are broken down according tothedenoted K$_{p}$ magnitude limits
%(close to the SDSS r magnitude). 
%Most of the analysis in this paper concerns presumed stars brighter than 
%16th magnitude. 
The means of the 4\,kK and 5\,kK dwarfs and giants are broken out separately. 
%A key question
%in this paper is why the mean (FUV-NUV) color becomes more
%negative for the coolest stars represented (4-5\,kK).
}
\label{fndist}
\end{figure}

Smith \& Shiao (2011) have pointed out two peculiarities in the distribution 
of GALEX UV colors of Galactic stars. These peculiarities raise questions 
which we address through this paper. The first peculiarity was a separation 
into two clean peaks of the distribution of stars' (FUV-NUV) colors. 
Similar distributions are exhibited in Figure\,\ref{fndist}for stars in
the Kepler FOV. These
objects are point-like and have circular shapes according to the GALEX 
pipeline, but particularly for the fainter objects are not necessarily 
stars. Objects brighter than K$_p$ = 16.0, i.e., those which we are 
studying, have a high probability of being stars.  We excluded a very small
number of bright objects (NUV, FUV $<$ 10 mag.) for which the pipeline 
flagged saturation in the recorded NUV or FUV pixels. Likewise, we
excluded stars for which the KIS pipeline flagged saturation.

% FIGGURE 1 -1 column figure: add * after "figure" at begin{figure},end{figure}
% for 2 column figures add * after "figure" - same rule for table

The presence of the two discrete peaks in Fig.\,\ref{fndist} is 
unexplored and is the first focus of our work.  
We noticed first that the relative heights of the two peaks differ
markedly for the faint (extragalactic) objects, Galactic dwarfs, 
and Galactic giants/supergiants - we define below our criteria for
segregating stars into dwarfs and giants/supergiants categories.
The galaxies have a strong UV-color peak.
We also compared the bimodal distributions of dwarfs and giants stars. 
Although the statistics are poorer, there are 
more dwarfs in the blue UV peak than giants.
%The fact that the dwarfs show 
%a markedly strong blue peak compared to the giants, despite their  
%being the apparently fainter group, suggests that no strong bias 
%exists against FUV magnitudes for UV-normal dwarfs in our sample.

\begin{figure}[ht!]
\centerline{
\includegraphics[scale=.40,angle=00]{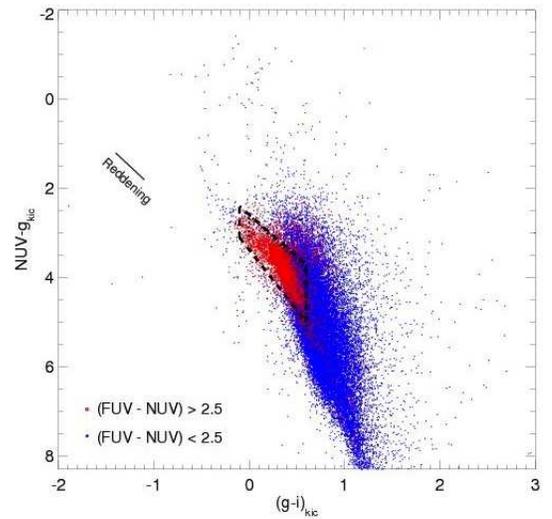}}%screenshot pdf
\caption{The $(g-i)_{kic}$, NUV-g$_{kic}$ diagram for K$_p$ $<$ 18 stars 
in the Kepler field. 
%All magnitudes NUV are given in the AB system.
Blue and red coded stars are differentiated by their (FUV-NUV) color, $\le$2.5
or $>$2.5. according to Fig.\,\ref{fndist}.
The dashed trapezoid is the region of the diagram taken from the UV-red 
stars observed {\it in the GALEX-SDSS sky} (Smith \& Shiao 2011).
The reddening vector represents E$_(g-i)$ = 2$\times$E$_{(B-V)}$.
% = 0.4 mags.
%A second question of this paper is the identity of the ``UV-red clump") stars. 
}
\label{ging1}
\end{figure}

  Next, we found it instructive to segregate members from these peaks
into ``UV-blue" and ``UV-red" populations. We have used the color criterion
(FUV-NUV) = 2.5, the midpoint between the peaks, for this separation, 
where FUV and NUV are determined in Oke \& Gunn's (1983) AB magnitude system. 
Next, we plotted colors in the 
% In plotting colors of our sample in the 
Bianchi et al. (2007) diagnostic $(g-i)$, $(NUV-g)$ diagram, 
in which the authors give useful diagnostics for stars. 
This plot, shown in Figure\,\ref{ging1} shows a 
second peculiarity, namely a relatively small ``UV-red island."  We will 
refer to this region in the context of this figure as the ``UV-red clump."  
This leads to a secondary focus of this work, which we can 
distill into the two related questions: what are the UV-red clump stars, 
and why are they surrounded by UV-blue stars, particularly by cooler 
stars in the lower right of this diagram? 
 
% FIGGURE 2 -  ging1

 All objects in our sample are flagged as point-like in the surveys we used, 
and because they are relatively bright, they are all stars, or in rare cases 
possibly QSOs. It then becomes important to separate subpopulations into 
KIC-derived effective temperature bins, corresponding approximately to spectral 
types, and luminosity class, in order to address the reasons for this 
peculiarity. Interestingly, we found that the median (FUV-NUV) colors of these 
bins do not continually increase monotonically with decreasing \teff~ (see 
Fig.\,\ref{fndist}), as would be expected, but rather attain a maximum at 
T$_{\rm eff}$ $\sim$ 6\,kK. 
This curious feature is called out by the annotations 
in the first figure, and it gives hints as to the reason for the presence of 
the red UV clump. As we will show in $\S$\ref{reslts}, this discovery links 
the two sets of questions above. Our work will investigate the reason for
the reversal of the GALEX UV color towards cooler effective temperatures.

\section{Data Analysis }
\label{anproc}

\subsection{General procedures}
\label{genproc}

  Although our study ultimately concerns stellar populations, 
especially in the Kepler FOV,  the fact
that the \teff~ values in the Kepler Input Catalog (KIC)\footnote{KIC, KIS,
and SDSS/DR7 photometric 
data are available for prospective targets in the Kepler FOV 
at the website {\it http://archive.stsci.edu/kepler} administered by the 
Mikulski Archive for Space Telescopes (MAST). Support for archiving MAST 
data is provided by NASA Grant NAS5-7584. These data are used ``as is," that is
without dereddening and without tranformations to a common magnitude system.}

are determined from photometric colors makes them less reliable than
if they were derived from spectra.
%particularly for the study of ill-understood photometric features.  
Instead we sought \teff~ values elsewhere for large numbers of 
faint stars that could be applied to stars in the FOV.  

   Spectroscopically derived ~\teff~ values for many faint stars from
the SDSS survey, and in the GALEX sky, can be obtained from the 
%   The requirement of spectroscopically-derived ~\teff~ values for stars 
% included in faint optical and also UV surveys can be met, thanks to the 
% overlap of sky regions covered by GALEX and SDSS, by obtaining the 
% T$_{\rm eff}$ values computed by the 
the SDSS grism pipeline (Vanden Berk et al. 2001,
Stoughton et al. 2002, Allende Prieto et al. 2008).
This pipeline processes spectra in the wavelength range 3800-9200\,\AA.~
The specra are delivered as a supplemental product for a small subset of 
targets that are also observed spectroscopically.
A technical summary of the spectrographs is given by Smee et al. (2013) 
The resolving power is R = 1850 in the blue and 2200 in the red. 
% corresponding to a resolution of 2-5\,\AA.~ 
The automated spectroscopic SDSS/SSPP (SEGUE Stellar Parameter Pipeline; Lee
et al. 2008)
subtracts a fitted polynomial from the
spectrum, leaving a difference spectrum of emission and/or absorption lines. 
The population type of point-like object is assessed from the morphology 
of the spectrum: 
a strong blue continuum and redshifted emission lines for QSOs, a red
continuum with redshifted broad absorptions/emission features for galaxies,
and unshifted absorption lines for stars.
Once a spectrum is evaluated as stellar the \teff, 
%\logg,~ and metallicity ([Fe/H]) parameters 
is computed and stored in the SDSS {\it sppParams} table as the parameter 
``teffa." We extracted these temperatures from the SDSS archive at MAST. 
These temperatures were used to bin spectra to the nearest kiloKelvin in
Figs.\,\ref{fnteff}. 
Apart from this figure, effective temperatures referenced for stars in 
the Kepler field are taken from the KIC.

Because we wanted also to check on the reliability of the 
GALEX photometric colors by means of this project's grism spectra, we
crossmatched all stars observed spectroscopically by both GALEX (GR6/DR7) 
and SDSS (DR7) over the whole sky. This netted spectra for 369
objects flagged by both missions as stars.
%for which the 
%NUV, $g$, $r$, and $i$ magnitudes had errors of $\le$${\pm 0.1}$ mags.;
%for FUV this limit was ${\pm 0.15}$ mags.
% and the brightest star has an equivalent K$_p$ = 13.76. 
%We also restricted ourselves to objects for which both GALEX and
%SDSS surveys have assessed to be point-like.
The GALEX grism observations that included these stars were made as part 
of a variety of GALEX surveys as well as Guest Investigator programs.
The \teff~ values computed by the SDSS DR7/grism pipeline 
lie in the ranges 4213-9445\,K, respectively.\footnote{A 
crossmatching of the 369 stars in common to the SDSS and GALEX surveys shows
that the \teff~ values from the more recent DR10/SSPP pipeline below 9000\,K 
exhibit an tyical offsets of -25\,K to -60\,K relative to the DR7 and an 
r.m.s. of 
${\pm 83}$\,K. To date the body of work done on \teff-color calibrations 
of late type stars for Kepler stars has not been replicated
for the DR10 temperatures. For this reason we use DR7 \teff~ values in
this work.}
%and 2.8-4.6 dex, respectively.
We did not attempt to evaluate the pipeline metallicities.
These \teff~ values ultimately formed the basis for the empirical 
calibration of \teff~ for all but the extremes of the stellar photometric
sample in this paper.  Of our joined GALEX-SDSS spectroscopic sample, all
but 4 stars have K$_p$ $>$ 14.0, and likewise nearly all have SDSS-derived 
\logg~ values consistent with their being dwarfs. 
%(we include the turn off 
%and subgiant subpopulations in the general dwarf description).
%values, which qualify them as giants. Since the great majority of these
%stars are presumed dwarfs, 
From this we can refer to their \teff~ values by their main sequence 
spectral types, and the above range 
corresponds to spectral types A1-K6 (Drilling \& Landolt (2000).
Our 1\,kK temperature bins have the
following spectral type equivalence: 9\,kK for spectral types for A3 and
earlier, 8\,kK for A4-A9, 7\,kK for F0-F6, 6\,kK for F7-G5, 5\,kK for
G6-K4, and 4\,kK as K5 or later. We designate these ranges as types 
BA, A, F, G, K, and late-K. 
%For the most part, these are 
%the spectral type ranges we can discuss without extrapolation. 
A few percent of our sample extends to early M dwarf and/or giant stars.
%objects that the Kepler Working Group on Stellar Properties have 
%classified as early M stars on the basis of more extensive spectroscopy, 
%IR photometry, and asteroseismological analysis.

  The procedure we follow is to calibrate the main sequence \teff~
distribution against the (FUV-NUV) color, and from that relation the 
SDSS-\teff~ with $(g-i)$ color. Once a ~\teff,~ $(g-i)$ relation is derived
from the common GALEX-SDSS spectroscopic sample, we applied it to KIC stars 
in the FOV observed by GALEX and the KIS survey in the FOV.
The GALEX colors for these are the same, but for the FOV region the 
use the $(g-i)_{kis}$ colors are expressed in the Vega system and 
therefore differ by +0.47 magnitudes (Bianchi 2011) from the 
$(g-i)$ SDSS survey (which is based on the AB magnitude system). 
%The constant 0.47 mags. was derived by Bianchi et al. (2011) 
%from the spectrumofVega itself and happens
%also to be in excellent agreement with the value
%+0.46${\pm 0.02}$ magnitudes reported by MAST from an empirical comparison
%of the color difference for the Vega and AB magnitude systems.
% this is the same 
%value found by MAST staff from a larger object sample (see
%http://archive.stsci.edu/kepler/kepler\_fov/explanations.html).
%%transformation of Kurucz atmosphere fluxes through these photometric 
%%filters; see http://dls.physics.ucdavis.edu/, site of the Deep Lens 
%%Survey).}
%The difference is due mainly to the necessary transformation of the $i$ 
%filter magnitude from the AB to Johnson system (or vice versa).
%%(which is on the AB magnitude system; Oke \& Gunn 1983) to 
%%the $i_{kis}$ (Vega-Johnson system; Johnson \& Morgan 1953). 
%%See the MAST site
%%{\it http://archive.stsci.edu/kepler\textunderscore fov/explanations.html}
%%for more details. 
%%GALEX %observations with the Kepler FOV, 
For purposes of comparison the mean reddening in the overall SDSS/GALEX field
we surveyed is E$(B-V)$ $\approx$0.05 magnitudes. This is a very modest amount
and is about three times smaller than toward stars in the Kepler FOV. 
In both surveys the reddenings are determined from an IR emission map over
the sky produced by Schlegel et al. (1998).
Because mean and the spread in the Interstellar Medium (ISM) reddening 
parameter given in the KIC is small, we did not attempt
to correct the colors to their unreddened values for intergroup comparisons.
%We note also that there is reason to believe from WISE photometry 
%(Wright et al. 2010),
%combined with spectroscopy and asteroseismology, that the KIC overestimates 
%reddening, especially for late-type giants (Pinsonneault 2013).
%This is due to an overestimate of the \logg~ values 
%for giants and also because 
In figures to be discussed, we display the computed reddening
vectors from various reported transformations (e.g., Straizys 
et al. 1998, Bianchi 2011). 
%In Fig.\,\ref{fngi} we omit the vector because 
%(FUV-NUV) is affected very little by reddening (Bianchi 2011).

\subsection{Selection of stellar parameters}
\label{selct}

\subsubsection{Effective temperature scale}
\label{teffs}

% We have adopted the SDSS \teff~ Stoughton et al. (2002) calibration 
%% (relying on grism spectroscopic and line aggregate indices)
%when it was advantageous to bin groups of temperature by temperature.
%Although at this writing the definitive corrected temperature scale 
%for the KIC is still being settled, we have corrected our calibration 
%to the Pinsonneault et al. (2012) scale, which is about 200\,K cooler 
%than the KIC calibration (T. Brown et al. 2011) 

  Our analysis depends on the calibration of SDSS and GALEX 
colors with \teff~ and our calibration is based on 
stars in the SDSS/GALEX sky.  Their temperatures have a 
rms errors of ${\pm 100-125}$\,K and are on average
-90\,K below the Pinsonneault et al. (2012) recalibration and about  
125\,K higher than the original KIC calibration for cool (\teff~ 
$\le$ 6.5\,kK) stars. In their photometric analysis of 
cool stars, An et al. (2013) adopted a 
\teff~ scale that is similarly 100\,K higher 
than the SDSS pipeline scale. 
The Kepler Working Group on Stellar Properties (WGSP) has likewise
recommended that a \teff~ scale close to the Pinsonneault 
one be used for the interim (Huber et al. 2014). \\
%The revised WGSP \teff~ scale extends over the range 12-3.2\,kK 
%(for main sequence stars B8-M1; for luminosity class III and I the 
%lower boundary is M1 and K7, respectively).  
%%Our results are not sensitive to these differences and are relevant 
%%only in binning of the \teff~ values into their 1\,kK bins.

\subsubsection{Surface gravity and dwarf/giant dichotomy}
\label{lgg}

Another parameter relevant to an understanding of anomalous color 
populations is the surface gravity. At best, we can hope for no more than 
a clean separation between ``giants" (luminosity classes I-III) 
and dwarfs. As we will see, the separation of these classes from high 
gravity objects (WDs, sdBs) can be confirmed by differences in 
red fluxes of the high gravity stars in various color-color diagrams. 
Primarily, the dwarf/giant separation can be accomplished by: 
a) evaluating the distributions of the two 
groups' apparent magnitudes (distances from Sun), 
b) determining \logg\, from photometry or spectrophotometry, e.g., 
by trusting values given in the KIC or (for non-FOV sky regions) the SDSS 
pipeline and rough determinations of \logg, and 
c) relying on photometric measurements of the size of the Balmer jump,
at least for the hot end (AF types) of our sample. 
%In the end, we might hope to match any relations we determine with 
%spectroscopic determinations of \logg~ from a subsample of stars in the 
%Kepler FOV. 
In our study we make use of all these diagnostics.
Most F-K stars in the four associations in the FOV (NGC\,6866, NGC\,6811, 
NGC\,6819, and NGC\,6791) are dwarfs and in principle would be counted as such.
Yet, the first three clusters were not included in the GALEX surveys,
and only one GALEX tile even brushed the edge of NGC\,6791 field. 
%Therefore, we could not make any dwarf/giant adjustments for these stars. 
Therefore, we could not make any luminosity class reassignments for these 
stars. This is unfortunate as it would be helpful to include clusters stars
of known ages.

% NOTE: Huber's slide 4 of 6/18/13 shows a H-K vs J-H diagram. All stars 
% with J-H > 0.74 are assumed to be a  giant (all sp types).

 Histograms of \logg's from our separate KIC (FOV) and the 
SDSS/GALEX populations exhibit similar bimodal distributions. 
A primary peak is centered at \logg = 4.0, and a secondary one at 3.2,
with an interpeak midway in between. 
Ciardi et al. (2011) classify as giants those F and earlier-type stars
with \logg $\le$ 3.5, G-type stars with \logg $<$ 3.7 as well
as K-M type stars with \logg $<$ 4.0. These criteria are 
consistent with the study of Galactic thick disk (r $<$ 4\,kpc) 
stars using SDSS spectra (Carollo et al. 2010). For those KIC stars 
examined by the Kepler WGSP and brighter than K$_p$ = 14, the
distribution of \logg's suggests that the percentage 
of giants and supergiants among AF stars is about 35\% and
includes most horizontal branch and RR Lyrae stars. 
The percentage of giants to dwarf levels off at $\approx$40\% for G stars
(Batalha et al. 2010) and rises again to about 50\% among K and early
M stars when the red giant branch is included (e.g., Mann et al. 2012). 
%As for middle and late-M stars (not investigated herein), the 
%percentage of M main sequence stars again becomes a dominant 
%contribution, largely because of a concentrated effort by the Project 
%to include M dwarfs in their sample of prospective exoplanet hosts.
%All these numbers are approximate since the WGSP estimates that even
%their improved \logg~ values have the same mean uncertainty, ${\pm 0.4}$ 
%dex, as the KIC advertises for its values.
The general conclusion is that the number of
dwarfs exceeds that of giants but by less than a factor of two. 
Aside from our exclusion, otherwise important, of faint (K$_p$ $>$ 16.0) and 
thus most mid to late M-type stars, we found no reason to believe that our
sample is different than the full KIC population.

Our provisional strategy was to use the simple Kepler magnitude
for the GALEX-SDSS star populations. 
We note here that the K$_p$ and $r_{kis}$ magnitudes are generally within 
${\pm 0.15}$ magnitudes of one another. We used the criterion described 
by Ciardi et al. 2011) as the demarcation between the space 
distribution of giants (K$_{p}$$\le$14.0) and dwarfs (K$_{p}$$>$14.0).
%, which form
%two distinct peaks separated by nearly 4 magnitudes in their populations
%as well as the ones we will investigate.
Of course this rough criterion admits some spillover from one group to the 
other. Mann et al. (2013) found that $\sim$7$\pm{4}$\% of K$_{p}$$<$14.0)
dwarfs fall into the so-defined giant domain. A somewhat smaller percentage 
of apparently faint giants, perhaps 4${\pm 1}$\%, fall into the dwarf pool. 
Therefore, as a secondary criterion, we required that the published 
KIC \logg~ must be less than 4.0 to 3.5 dex, according to the Ciardi et al. 
demarcations.  In the occasional conflict between these two criteria 
or if the KIC listed no \logg\, value, we classified the star as a dwarf. 
%In addition to these factors, we excluded stars that are close to the
%centers of the open clusters NGC\,6811 and NGC\,6819, as specified by
%Meibom et al. (2011) and Balona et al. (2013), respectively, which would 
%otherwise include a number of apparently bright dwarfs near their main 
%sequence turnoffs; according to the K$_{p}$$>$14.0 demarcation they would
%automatically be counted as giants.
% As a consistency check, we compared the KIC published \logg~ values 
% of 220 stars known as Kepler Objects of Interest (Batalha et al. 
% 2010) and \logg~ values rederived spectroscopically by Everett et al. 
% (2013).  Some 81\% of their stars satisfy the K$_p$$>$ 14 criterion,
% and of the total sample their \logg~ values are about 0.15 dex smaller
% than the KIC values. The KOIs are biased toward bright stars, so between
% this selection effect and the KIC's low \logg values the actual success
% rate of this criterion is somewhat higher than 81\%.
This agrees with the findings of Huber et al. 
catalog (2014), which used spectroscopy
and asteroseismology to estimate \logg\, values. Given their
study, 23\% of ``unclassified stars" (those 7\% of KIC stars with K$_p$ $\ge$ 
14 and without listed \logg\, values) are giants. 
%Moreover, about 3\% of 
%the bright stars we have classified as dwarfs can be expected to be giants.
In terms of ``UV excess stars" described below, 98\% 
of this group that we classify as dwarf stars are dwarfs (log\,g $\le$
4.0) in the Huber et al. catalog. Conversely, 96\,\% of the UV excess stars
we call giant stars are giants in this catalog.
% Here are the stats (11/15/13):
% Of 470 UVe dwfs in 'binary' KIC list, 295 match Huber catalog
% Of these 288/295 have logg > 4.0, i.e. are dwarfs - 98%.`
% Of 88 UVe gnts in 'binary' KIC list, 71 match Huber catalog
% Of these 68/71 have logg < 4.0, i.e. are giants - 96%.
% %age of stars observed by Kplr: (295.+71)/(470+88) = 66%
%This is much somewhat lower than the 93\% estimate inferred from Mann et al. 
%The vast majority 
%%($>$90\%) 
%can be potentially verified as true dwarfs according to a number of 
%spectroscopic or asteroseismological studies (e.g. Huber 2014).
Although the dwarf/giant dichotomy enters into our discussion below,
our conclusions are not sensitive to the sizes of the two populations.

The study of stellar ages for solar-like stars in the thin Galactic disk 
%, for the most part stars studied in this paper, 
has meanwhile proceeded apace. The derived ages support the
general mix of giants and dwarfs in the KIC catalog.  
Although nucleocosmochronological 
techniques applied to the oldest stars in the thin disk suggest that some
of them formed 8\,Gyr ago (del Peloso et al. 2005), 
recent studies of the various stellar populations 
of the Galaxy (e.g., Soderblom, Duncan, \& Johnson 1991, Carollo et al. 2010)
generally find a typical age of 3-4\,Gyr for stars in the solar neighborhood. 
%In addition, Chaplin et al. (2011) have put into an evolutionary context 
%the distributions of basic stellar parameters (KIC \teff, radius, mass,
%luminosity) of 500 local stars in the KIC catalog and monitored by the Kepler 
%spacecraft with measures from solar-like oscillation criterion for stars 
%in the range 7\,kK $\gtrsim$ \teff $\gtrsim$ 5\,kK. 
%For reference they find masses for these stars to fall in the range 
%1.0-1.9M$_{\odot}$ and, except for supergiants, 1-3R$_{\odot}$. 
%In addition, according to the dwarf/giant separation criteria we adopted,
% these figures suggest 
%the population we study has a 60:40 mix of dwarfs and giants.  This is 
%also the ration found by Batalha et al.
%The isochrones of their superimposed evolutionary tracks suggest
%ages from essentially the zero age main sequence to 12\,Gyr (e.g., 
% Yi et al.  2001). 
In terms of stellar subpopulations this range includes F-K 
main sequence stars, BAF horizontal branch stars (including RR Lyr and 
$\delta$\,Sct variables) and GK giants. It is chiefly to these groups
that we now confine our attention.
%(As with our own sample, late K and M stars were not included, so this 
%is not representative of the true Galactic distribution.)  

\section{Results}
\label{reslts}
\subsection{GALEX and SDSS photometry and spectroscopy}
\subsubsection{UV-optical colors and stellar parameters }
\label{glxcol}

%\noindent{\it General}

\begin{figure}[ht!]
\centerline{
\includegraphics[scale=.40,angle=90]{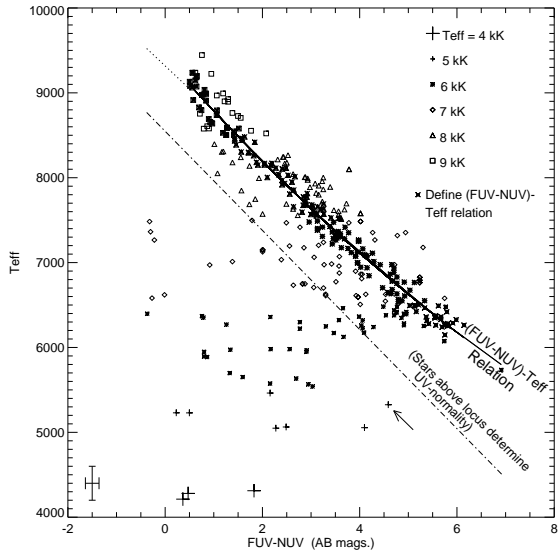}}
\caption{
(FUV-NUV) versus SDSS/DR7 \teff~ values derived by 
the SDSS grism pipeline for objects flagged as stars in the 
GALEX-SDSS sky and also having GALEX spectra. Symbols denote 1\,kK
 bins in \teff.~ 
%which the stars have been grouped. 
Solid symbols denote the stars used to define the upper (FUV-NUV)-\teff~ 
diagonal sequence and calibrate this color from their \teff~ values. 
% No reddening correction has applied to the stars' colors.
% (the mean computed E(g-i) $\approx$ 0.29. 
The dot-dashed locus separates the calibration sequence
from the ``UVe" population in the lower left.  The star highlighted by
an arrow is discussed in the text.
%at (FUV-NUV) = 4.5 is discussed in the text.
%indicates the position of one star with only a mild UVe color for its 
%optically derived \teff~  and for which the UV and optical spectra are 
%displayed in Figs.\,6, 7, \& 8. 
Note that cool stars (T$_{\rm eff}$ $<$ 5kK) with weak UV fluxes are 
not represented in our magnitude limited sample.}
\label{fnteff}
\end{figure}

  Bianchi et al. (2007, 2011) used the $(g-r)$, (FUV-NUV) color diagram with 
model Spectral Energy Distributions (SEDs) to differentiate stars from 
extragalactic objects, and to establish a stellar \teff~ relation with 
(FUV-NUV). These authors pointed out this color is nearly independent of 
reddening for a normal Galactic ISM color law
%because the central wavelengths of
%the FUV and NUV bandpass straddle the wavelength of maximum extinction 
(see Table 2 of Bianchi 2011).  We plotted 
T$_{eff}$ values determined from the SDSS spectroscopic pipeline 
for the stars that SDSS and GALEX observed spectroscopically.
The first result of this exercise is a diagonal sequence in 
Fig.\,\ref{fnteff} that reflects the relationship between SDSS \teff~ 
and (FUV-NUV). 
%A second and surprising result is the existence of a broadly scattered 
%distribution in the lower left region of the diagram. 
%All points are shown by symbols denoting \teff~ rounded to the closest
%integral kiloKelvin. Thus, stars with \teff~ in the range 8500-9500\,K
%are depicted with square symbols for ``9\,kK," etc.

We identified a group of stars within ${\pm 250}$\,K of a smoothed quadratic 
fit of the diagonal distribution (solid symbols). These were mapped
to the $(g-i)$, (FUV-NUV) plane and a quadratic fit was then computed 
to define a calibration sequence for UV-normal stars. 
We next transformed this sequence by adding 0.47 magnitudes to the 
$(g-i)$$_{kic}$'s to obtain the equivalent in the KIS (Vega) system.
Note here that many faint K and M stars with normal, weak UV fluxes are not 
captured in this sample.

% FIGGURE 3 - fnteff

% FIGGURE 4 -   fngi
\begin{figure*}  [ht!]
\centerline{
\includegraphics[scale=.90,angle=90]{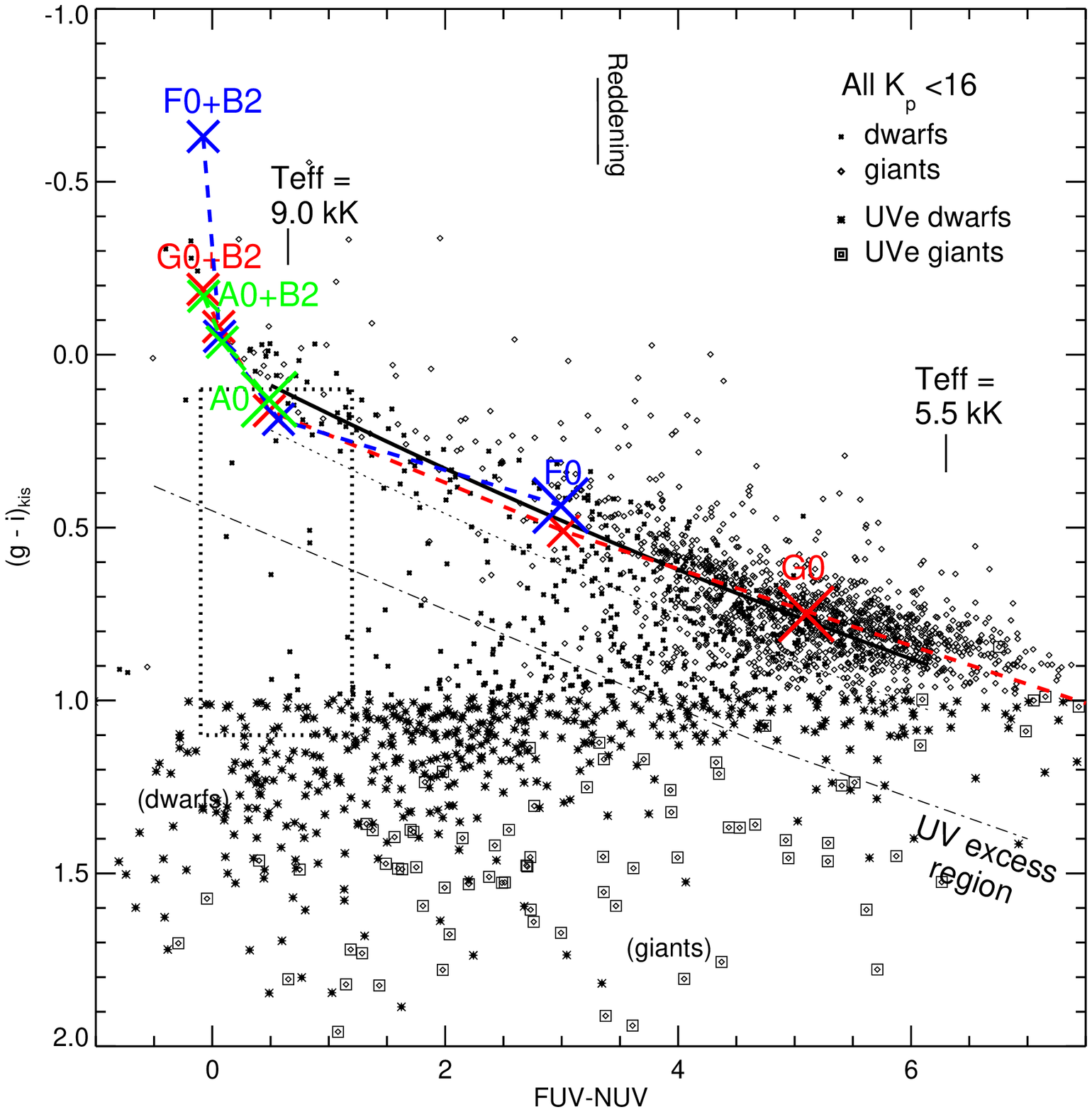}}
\vspace*{-.15in}
\caption{(FUV-NUV) colors plotted against $(g-i)$$_{kis}$ for 2285 Kepler
FOV stars, i.e,. those brighter than K$_p$ = 16. 
%and having SDSS spectroscopic solutions for T$_{\rm eff}$.
The dotted box gives the region occupied by most
extragalactic sources (Bianchi et al. 2007).
The solid line is computed from applying the relation shown in 
Fig.\,\ref{fnteff} to $(g-i)$$_{kis}$ colors. 
% and transforming it to the KIS (Vega magnitude) system.
%; a correction of this line to
%unreddened colors is made from the mean E(B-V)'s given in the
%directions of these stars from the KIC catalog. 
Annotations show the mean (FUV-NUV)'s of \teff = 9.0\,kK and 
5.0\,kK stars.  UV-normal and UVe stars are those occurring below
the dot-dashed line, whose position is likewise taken from an
estimate in Fig.\,\ref{fnteff} of excessive deviations ($>$3$\sigma$)
from the mean \teff~ calibration. Dwarf and giant UVe stars are 
denoted open squares and large asterisks.  
Colored lines and X's are modeled binary colors. 
%assuming the binary
%to be composed of pairs of main sequence stars computed from the same models. 
The computations include mean reddening and represent binary pairs
A0+B2, A0+B6, both A0 (no blueing) for an A0 primary; blue line: the
F0 sequence F0+B2, F0+B6, F0+A0, and both F0; red line: G0 sequence
G0+B2, G0+B6, G0+A0, G0+F0, and G0-G0.
% The (FUV-NUV) and $(g-i)_{kis}$ colors are in the AB and Vega systems, respectively. 
Like-component binaries are shown as large X's.
}
\label{fngi}
\vspace{0.15in}
\end{figure*}

Figure\,\ref{fngi} exhibits these stars plotted in the  
%the result of this transformation to 
(FUV-NUV) $(g-i)$$_{kis}$ plane and the UV-normal sequence curve.
%\footnote{Colors are taken 
% ``as is" from the MAST catalogs, {\it with} reddening and without 
% transformation to a common magnitude system.}
Note that by definition our sample of stars has
(FUV - NUV) colors and thus is magnitude limited. This means 
that some mid-K and nearly all M stars with red UV colors are excluded.
%Because the mean reddening in the FOV is small to moderate, our plotted 
%colors are ``out of the catalog,", i.e., with no reddening corrections applied.
% translations of the relations for E(B-V) = 0 and (with reddening) for 
%a different stellar (and by definition a magnitude limited) sample, 
% $(g-i)$ in the SDSS system.  
%These stars also have observed KIS $(g-i)$ and GALEX (FUV-NUV) colors. 
%For reference this diagram also shows the position of the 
%mean (FUV-NUV), $(g-i)_{sdss}$ locus if no reddening were present. 
%(Again, the $(g-i)$ colors used in the sample shown in Fig.\,\ref{fngi} 
%were obtained from the KIS release.)

Figs.\,\ref{fnteff} and \ref{fngi} exhibit a secondary scatter 
in their respective lower left regions, that is, points falling below 
the dot-dashed line. The shifted colors of this population 
are mainly due to a blueing of the (FUV-NUV) color of typically 3-4
and occasionally $\gtrsim$\,5 magnitudes
and a much smaller amount of reddening in the color $(g-i)_{kis}$. 
%These shifts are much larger than the predicted photometric errors. 
The displacements from the primary sequence are less on average 
for the hotter stars and greatest for the cooler ones.
We refer below to these objects as UV-excess (``UVe") stars.
We also note that when plotted on the sky plane the UVe stars 
exhibit no tendency to cluster, e.g., from membership in associations.
% Their spatial distribution is indistinguishable from the full population.
%%with GALEX photometry.
% About 17\% (62/369) of the stars represented in Fig.\,\ref{fnteff}
% reside to the left of a dividing (dot-dashed) line. 
%The percentage of stars in the lower left region of Fig.\,\ref{fngi} is
%nearly the same as this, 16\%, and seems therefore to be constant for both 
%the FOV (Fig.\,\ref{fngi} and the GALEX-SDSS sky (Fig.\,\ref{fnteff}).

All but two stars represented in the lower left region of Fig.\,\ref{fnteff} 
have K$_p$ $>$ 14.0, and all but eight have \logg$_{kic}$ $\ge$ 3.5. 
This suggests that the great majority of them are dwarfs.  As for the stars 
of Fig.\,\ref{fngi} occupying the same region of the diagram, only about 
15\% (79/506) are giants, according to our brightness and KIC \logg~ criteria.
% 15\% (86/506) are giants, according to our brightness and KIC \logg~ criteria.
%7\% (33/508) are giants, according to our brightness and KIC \logg~ criteria.
%about 25\% (200/786) are giants, according to our brightness and KIC \logg~ 
% This is probably a high estimate as the lower estimate given
% above from the GALEX-SDSS sample, despite its smaller populations, is 
% probably more accurate because the \logg~ values from the SDSS are
% more precise. 
Also, one finds there that the 
ratio of UVe giants to dwarfs increases as one proceeds to the cooler K stars.
% - probably because the K stars are dominated by red giants
In short, 
%17$\pm{1}$\% of stars represented in this figure fall in the lower-left 
22$\pm{1}$\% of stars represented in this sample 
fall in the UVe zone to the lower-left.
(The $\pm{1}$\% error is derived by comparing the statistics in two halves
of the FOV covered by the two KIS surveys.) This fraction is similar to 
the 17\% figure we found in Fig.\,\ref{fnteff}. 
Thus, we see that the statistics of the UVe stars are similar 
%(although possibly not identical) 
for populations observed using KIS photometry in 
the FOV and the high Galactic latitude portion of the sky 
covered by the SDSS survey. 
% Note that any difference in mean reddening of the stars in the
% Kepler and GALEX-SDSS regions of the sky is compensated by our tying
% the demarcation line to the stars defining the Teff-(g-i) calibration.

%In all the most prevalent single subpopulation of stars represented
%are G-K stars that are partially evolved off the main sequence, but
%not yet to the giant branch.

   From the annotations in Fig.\,\ref{fndist} we can also see the 
effect of the differing degrees of scatter in the (FUV-NUV) color 
for stars in various \teff\, bins. 
Fig.\,\ref{fndist} shows that the stars with the greatest UV blueing
are the coolest ones.
%, particularly those in \teff~bins 4\,kK, 5\,kK, and even 6\,kK. 
Also, the incidence of UVe is smallest among hot stars.
Interestingly, when the giants are split out from the dwarfs, 
we see that the 5\,kK dwarf group has a much bluer $<$$FUV-NUV$$>$ than
the 5\,kK giants. To a smaller extent this is true for the 4\,kK (early
M-type) stars as well. This is another way of demonstrating 
that the late-G/early-K dwarfs are subject to greater UV excesses
than the giants. When arriving at our few M stars, the (FUV-NUV) colors of 
UVe giants exhibit decreases by similar amounts as dwarfs. 

%Inspection of Figs.\,\ref{fnteff} and \ref{fngi} further suggests that 
%although the UVe zone includes some high \teff~ stars, they are relatively
%less numerous than their same-color cousins. 
%The conclusion here is that the UV blueing we have found is associated 
%mainly with stars having masses of a solar mass or less and for gravities
%consistent with moderate ages as they begin to move off the main sequence.
%However, the effect is still there for redder stars on the red giant 
%branch as well as early type stars near the main sequence. 

%  We note also that the mean UV color, $<$$FUV-NUV$$>$, is significantly 
%bluer for the dwarfs than the giants in the lower left zone (1.34${\pm 0.06}$ 
%versus 2.05${\pm 0.06}$). The mean $(g-i)$ color in the same zone is modestly 
%bluer for dwarfs than for giants (1.26${\pm 0.22}$ versus 
%1.51${\pm 0.35}$) - keeping in mind that the $(g-i)$ colors are 
%nonetheless redder than their upper right zone counterparts.
%%This is consistent with the notion that the UV light being added
%by a hot source, e.g. a blue secondary in a binary system, for which 
%the contribution is proportionally greater in the UV than in longer
%wavelengths. The effect is more evident for redder optical primaries
%in this paradigm. 

Our sample of GALEX-observed stars in the FOV imposes two countervailing 
biases. The first bias depends upon the ratio of sky areas observed in the 
NUV and FUV bands and second the loss of detections of FUV-normal (``red") 
stars. The area ratio is 1.15 and is magnitude independent. The second 
effect increases the number of faint UVe stars relative to UV normal ones. 
% stars due to observations only in the NUV band for stars too faint to be
% observed in FUV. This will decrease the number of 
%that discriminates moreso against UV-normal stars.
%Before continuing, we remark that our population is subject to a bias from
%being magnitude limited, and in particular suffers from an underrepresentation
%of stars observed by GALEX in the FUV band relative to the NUV. 
%The first component of this concerns sky areas observed only in NUV. 
%This bias amounts to a relative sky coverage of 
%about 15\% and is independent of the star's color. 
% is more depends upon the UV redness. 
At K$_{p}$ = 14.0 we find this bias in our reddest sample 
($(g-i)$$_{kis}$ $\ltsim$ 1.0) is a 15\% effect.
%and probably has its origin mainly in the sky areas observed in the two bands. 
As one goes to K$_p$ = 17.0
%fainter stars, for example, magnitudes but restricts attention to early-type 
%M stars with $(g-i)$$_{kis}$ $\ltsim$ 1.0, 
the bias increases to 40\%. Even though our study limit was K$_{p}$ = 16.0, 
we use the K$_p$ = 17 limit figure.
%and found the ratio to increase to
%1.40 for still fainter magnitudes.
%due to the relative paucity of FUV-observed stars. 
Then the ratio of these 
two values, 1.15/1.40 = 0.82, will be used a rough estimate of the 
incompleteness of the (FUV-NUV) colors and the fraction of 
faint UV-normal stars that could be missed by the GALEX surveys. 
Thus, we expect their UVe numbers are 
overrepresented and should be corrected by this factor. \\

\noindent{\it Do the UVe stars stand out among known binaries?}

\begin{figure}[ht!]
\centerline{
`\includegraphics[scale=.35,angle=90]{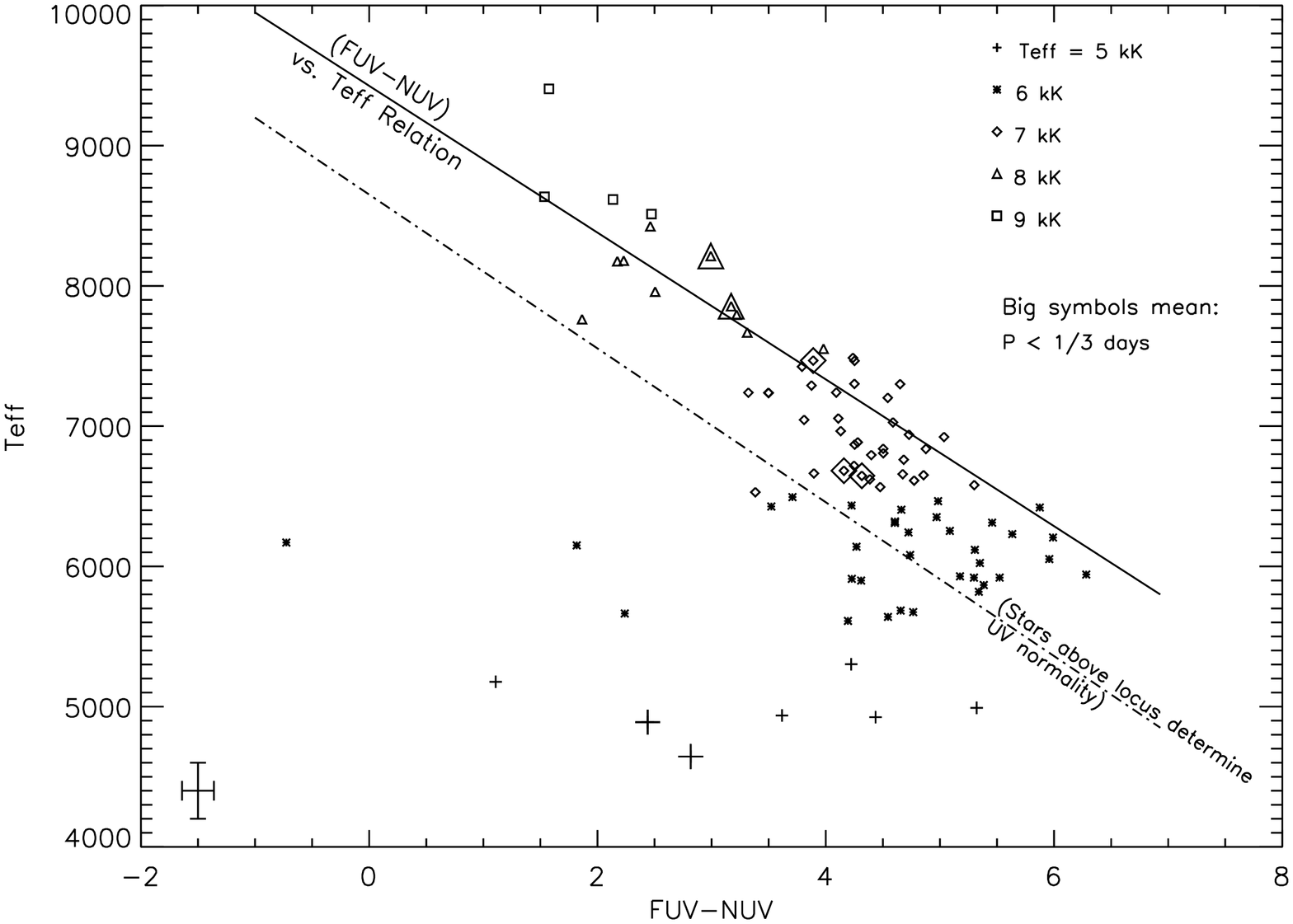}}
\caption{
The (FUV-NUV) colors plotted against KIC \teff\, values for eclipsing
binaries in the EBC which have also been observed by GALEX.  Symbols denote 
the same 1\,kK bins as in Fig.\,\ref{fnteff}, and the solid and 
dot-dashed lines are also taken from that figure. 
%bins in which the stars have been grouped. 
Large symbols indicate
systems for which P$_{orb}$ $<$ $\frac {1}{3}$ days. 
%Note that the
%distribution of points is similar to that exhibited in Fig.\ref{fnteff}.
}
\label{fnteffebc}
\end{figure}

  To pursue the possibility that the UVe stars are in binary
systems, we consulted the Kepler Eclipsing Binary Catalog (EBC), version
3\footnote{The active (in this case, third) version of this catalog may 
be found on line at http://keplerebs.villanova.edu.} 
(Matijevic et al. 2012)
%First, we cross-matched the stars with the 786 UVe stars from
%%First, we cross-matched the stars with the 451 UVe stars from
%Fig.\,\ref{fngi} and this netted 10 stars of all magnitudes up to 16.
%However, since the selection effects in this catalog are strong and
%certainly not all well known, we could not sensibly compare the 
%numbers of expected UVe stars from the EBC with those 
%found from our previous sample. Therefore, we reversed the question
and asked whether the stars listed there and also observed by GALEX 
show similar UV excesses.  We found 116 such stars. 
%of all magnitudes (all but one with K$_p$ $<$16).
The KIC \logg~ values for these stars range from 3.4 to 4.7. 
Even given the uncertainties in these photometrically derived values,
few members of this sample are likely to have 
optical primary giants or supergiants, and probably none are WDs.
Figure\,\ref{fnteffebc} depicts the relation between (FUV-NUV) and the KIC 
effective temperatures. One can see immediately that the distribution is 
much the same as in Fig.\,\ref{fnteff}. In particular, the number of UVe 
stars in the lower left of the figure is 20, corresponding to a fraction 
of 17$\pm{4}$\%, i.e., within the 17-22\% range found for the GALEX-SDSS 
and Kepler FOV samples. In addition, the hotter main sequence stars are 
again relatively rare, and the (FUV-NUV) blue shifts are the largest for
the coolest EBC primary stars.  We also find that the distribution of 
orbital periods of the UVe stars is consistent with the UV-normal 
star population. For example, if we take those few stars with 
P$_{orb}$ $<$ $\frac{1}{3}$ days, a value chosen to mitigate
effects in synchronous or mass-transferring binary systems, we still find 
no real difference between the UVe and UV-normal subsamples. 
%this subpopulation is represented 
%proportionately the same among the UVe and ``normal" populations. 

% FIGGURE 5 -

 We have also examined 12 members of the Kepler Objects of Interest (KOI),
as listed (as of July, 2013) in the MAST archives,
%\foootnote{Since this date
%new versions of the KOI list have primarily {\it removed} false positives 
%(eclipsing binaries), which of course is contrary to our current interest.}
that have available (FUV-NUV) and $(g-i)_{kis}$ colors and \teff\, 
or \logg\, values from the EBC. 
All 12 have F-G spectral types, and all 
but one are dwarfs. We find their distribution in the (FUV-NUV),
\teff\, diagram to be similar to those shown in the other figures.
Only 2 of the 12 KOI members, K03868.01 and K01345.01, fall in the UVe
region of this and the (FUV-NUV), $(g-i)_{kis}$ diagrams. 
This is consistent with the $\sim$20\% fractions found for our two 
larger populations.
%It remains to be added that for the KOI members the mean value 
%of $(r_{kis}-K)$ is about +1.0, or about the same as that found for 
%dwarfs ``binaries" in Fig.\,\ref{rkfn}.

  We see from this study that the UVe population has a
similar presence in EB systems and the general Galactic disk. 
%In other words, the phenomenon is consistent with binaries systems comprising 
%the population, with the primary component being the brighter in the visual and
%the secondary dominating in the UV. 
This does not mean that binarity is a necessary condition for the UV excesses.
% or that it is part of the cause 
%of the UVe phenomenon. 
In fact, as a counterexample to a binary scenario we can point out that one 
short-period Algol system in the Kepler FOV, WX\,Dra (KIC\,10581918), 
fits the physical description of what a UVe system might look like. 
The KIC (photometric) T$_{\rm eff}$ is 7252\,K for this system. 
%which is equivalent to an $\approx$F0 star, i.e., between the spectral
Spectral types for the two components have been estimated as A8 and KO 
(Budding et al. 2004).
However, the colors of this system fall on the normal side of the UV-normal/UVe
demarcation in Figs.\,\ref{fnteff} and \ref{fngi}. Clearly, binaries with 
cool and warm components are not necessarily UVe objects.
%In $\S$\ref{discs} we will rule out this possibility more completely 
%as well as the likelihood that old single stars can show UVe properties.
%%
%%Therefore, in the following we will refer to these UV-strong stars 
%%in the lower left zone occur as ``binaries," with the quotations 
%%retained until such time as their identity can be better established.
%%Also, it is now obvious that not all binaries with F-K primaries
%%have UV colors.

\subsubsection{GALEX/NUV grism spectroscopy }
\label{grsm}

   In one of the few papers devoted to the characteristics of GALEX spectra 
of normal stars, Bertone \& Chavez (2011) have demonstrated the common
sense expectation that the flux gradients of spectra obtained with the 
NUV camera correlate well with spectral types, at least for Henry Draper
stars in the range A0-K0. Surprisingly, no corresponding demonstration has 
been published to date for FUV spectra.

\begin{figure*}[ht!]
\centerline{
\includegraphics[scale=.75,angle=90]{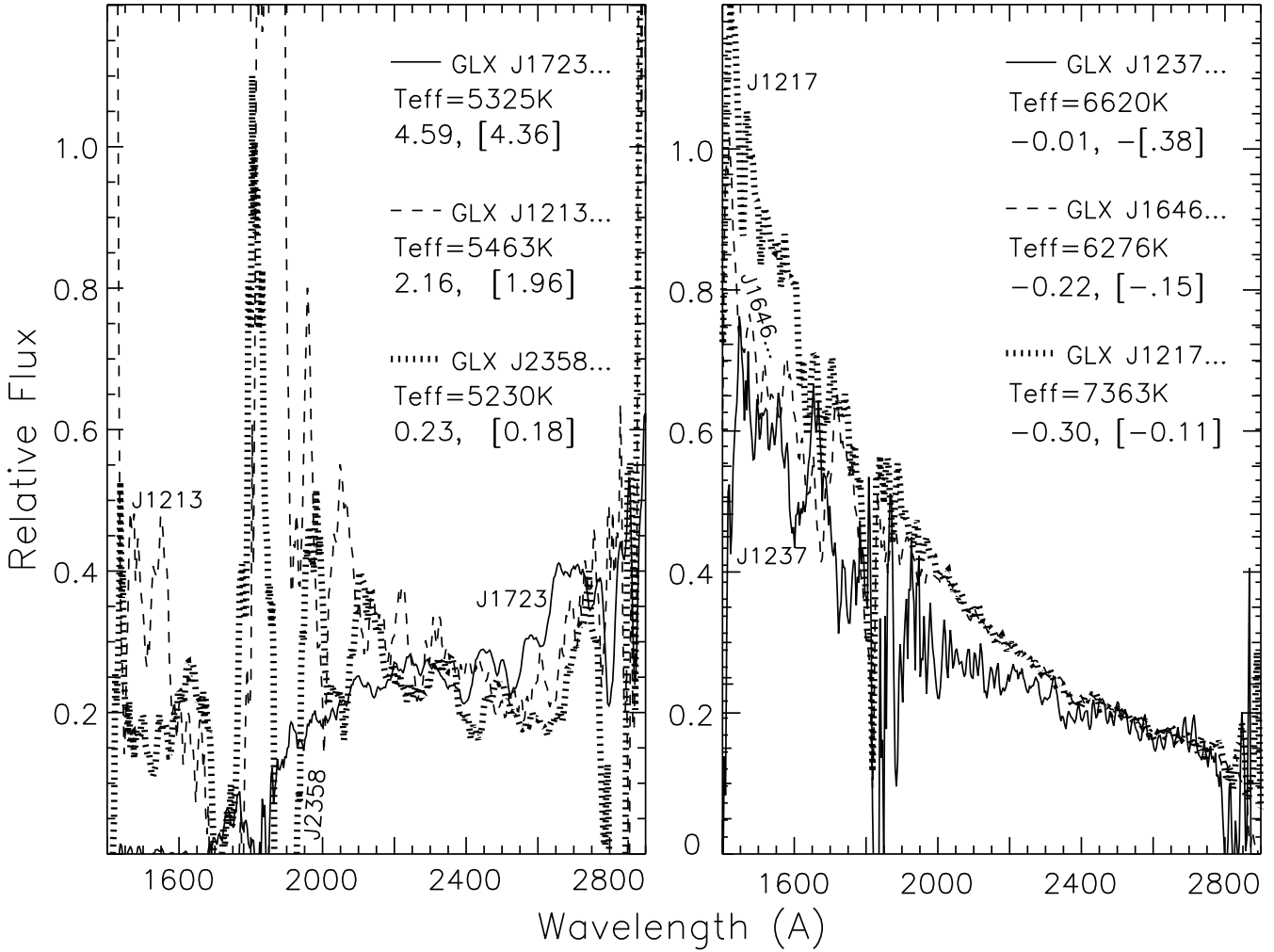}} % 2-pnl of grism spectra
 \vspace*{0.15in}
\caption{ GALEX grism spectra of six selected stars observed
photometrically and spectroscopy by both GALEX and SDSS.
%sky, which have been observed in both GALEX FUV and NUV cameras and also
%the SDSS grism spectrograph. 
Fluxes are smoothed by a running 4-point boxcar
function.  Annotations for spectra include name,  SDSS \teff,~ and (FUV-NUV)
colors (observed; computed colors from spectra in brackets).
Spectra are arranged in order of (FUV-NUV) color. 
The first spectrum in panel a (solid line), 
GLX J172312.5+593928, already a mild UVe star.
%has a red UV color, but it still suffers some UV blueing
% relative to other stars of the same temperature (see Fig.\,\ref{fnteff}). 
The wavelength range 1700-1900\,\AA\, lies at the edges of the FUV and NUV 
detectors; in panel a the fluxes are very low and represent noise. 
The negative spectral slopes in panel b 
% are consistent with the negative values of the UV color
and suggest a contribution of far-UV flux. 
%inconsistent with the star's optical spectrum. 
%Colors in brackets are our simulations run through GALEX filter curves.
%[Note that none of these spectra shows emission lines. \
}
\label{2pnl}
\end{figure*}

To rule out that the UVe stars' small and negative (FUV-NUV) colors 
can arise from noise sources in the FUV filter, and recalling that 
there is almost no overlap of objects observed in SDSS 
grism program in the Kepler FOV,
%(e.g., short wavelength noise spikes where the stellar flux is undetectable), 
we searched through all the SDSS spectroscopic catalog for objects 
observed in the common GALEX-SDSS sky. 
Examples of these matches are shown in Figure\,\ref{2pnl} and are ordered 
in the panels by increasing (FUV-NUV) from lower left to upper right. This
figure also annotates the SDSS \teff\,~ values and GALEX (FUV-NUV) colors.
Also, following Bianchi (2011), we list computed GALEX colors that we
computed by running spectral fluxes through GALEX 
filter transmission curves. 
These track both the catalog colors and the continua slopes. 
%In all, the progression of continuum slopes agrees well for the (FUV-NUV) color.
%According
%to one interpretation, this could suggest the contamination of the optically 
%bright star by a strong UV source. Alternatively, it would suggest an
%atmosphere with a chromosphere that extends into the photosphere. In any 
%case,

\begin{figure*}[ht!]
\centerline{
\includegraphics[scale=.77,angle=90]{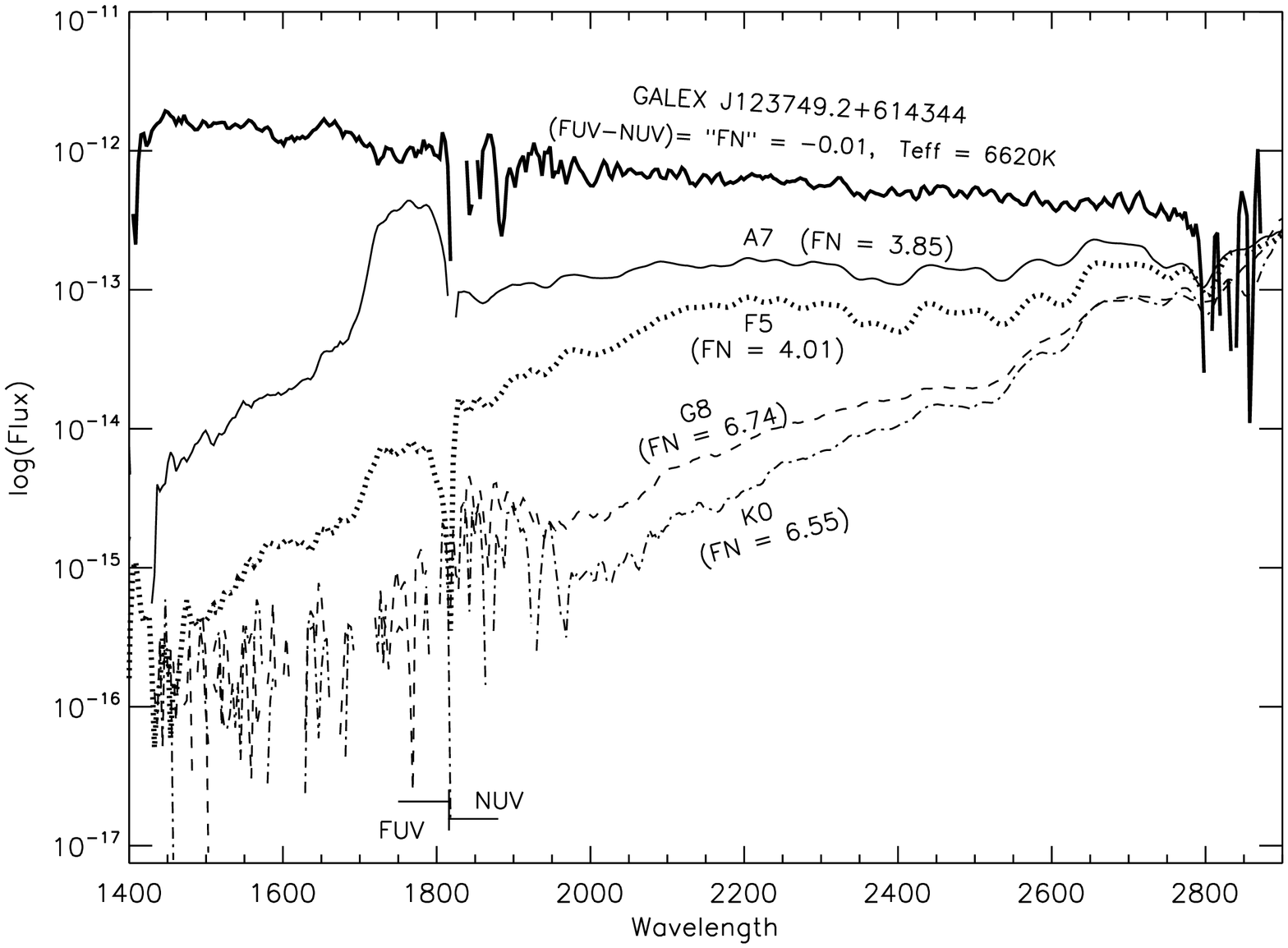}} 
\caption{ FUV+NUV GALEX spectra of four Henry Draper
quasi-standards discussed by Bertone \& Chavez. 
%(2011) discussed NUV spectra. 
Spectral types (annotated) are consistent with the differing slopes
of the spectra. The reference spectrum of GLX J123749.2+614344 (dark 
solid line) is also given in Fig.\,\ref{2pnl}.
The splice point for FUV and NUV spectra is indicated. The flux hump at 
1700-1800\,\AA\, for A7 is a noise artifact.
}
\label{bertn}
\end{figure*}

The solid line in Fig.\,\ref{2pnl}a depicts a GALEX spectrum of a G
star with a modest positive slope with increasing with wavelength.  
However, this UV continuum slope is inconsistent with the star
the star having a \teff = 5325\,K, according to its SDSS optical spectrum. 
This example is tagged as the point marked with an arrow in Fig.\,\ref{fnteff} 
- it is a mild member of the UVe population. 
%even though it is a reference standard in Fig.\,\ref{2pnl}.  
Indeed, the decrease in this star's flux 
from 2700\,\AA~ to 1900\,\AA~ is only a factor of two or so, whereas
according to Bertone \& Chavez the flux reduction across this range should 
be ten times steeper for G stars of this effective temperature, 5325\,K 
(see Figure\,\ref{bertn}).
% This emphasizes that the star is an (albeit mild) UVe member.

% FIGGURE 6 ; 2 panel grism spectra  -

\teff\, values of six F-G star examples in Fig.\,\ref{2pnl}
lie in the range 5230-7363\,K, and
their (FUV-NUV) catalog colors range from +4.59 to -0.30. 
%Four stars in this group were assigned \logg's of $\ge$4 by the 
%SDSS processing pipeline, which identifies them as dwarfs. 
%The remaining two examples happen are giants.
Two of the three spectra represented in the left panel have \teff~ values 
consistent with their being late G stars, 
from the SDSS pipeline whereas two of the examples in the right panel, 
associated with negative (FUV-NUV) colors, have \teff\, values corresponding 
to F types. 
%again for the optical primaries.  
This is consistent with the smaller blue displacements in UV color in 
Fig.\,\ref{fnteff} for warmer members of the UVe population and suggests 
once again that UV excesses are smaller for warm stars.
%that the displacements caused by secondary contamination for the 
%reference star are mitigated by the still measurable UV flux of an 
%already warm primary.
Note also that the spectra do not show even a hint 
of emission in the C\,IV or Mg\,II h/k doublets.
% this is also true of the unsmoothed spectra.

  Fig.\,\ref{bertn} assembles FUV/NUV-merged spectra
for four HD stars with types A7-K0 represented in Bertone \& Chavez (2011). 
Annotations show the GALEX catalog UV color. 
%our simulations (not shown)
%parallel the monotonic increase in color with type. 
The montage also compares
these spectra with a UVe spectrum (shown in bold) displayed also
in Fig.\,\ref{2pnl}. The UVe star represented has a SDSS (optical) spectrum 
consistent with a middle F star; the UV continuum slope is indicative
of an A star.

   In assembling UV colors of these standards we note a curiosity 
that is confirmed in the catalog and simulated GALEX colors:
the change in advancing color with spectral type ultimately 
decreases among the mid-G and K stars. 
In our G8 and K0 examples the UV color can overlap. 
The spectra show that this is because although the FUV
continuum decreases with late types the strengthening of the NUV lines
increases as well; the two effects nearly cancel. We speculate that this is 
because by $\sim$G5 
the NUV lines are still saturating whereas the lines in the FUV
have already depressed the spectrum.

% FIGGURE 7 ; Bertone + 'UVe' grism spectra  -  bertn

  In any case, the key point from the spectrum figures is that since the 
continuum slopes track the (FUV-NUV) colors of these stars, there is no reason 
to doubt the GALEX colors. Instead, we must look to explanations involving
stellar classes for the existence of UVe stars. 

\subsubsection{SDSS (optical) grism spectroscopy } 

\begin{figure*}[ht!]
\centerline{
\includegraphics[scale=0.77,angle=90]{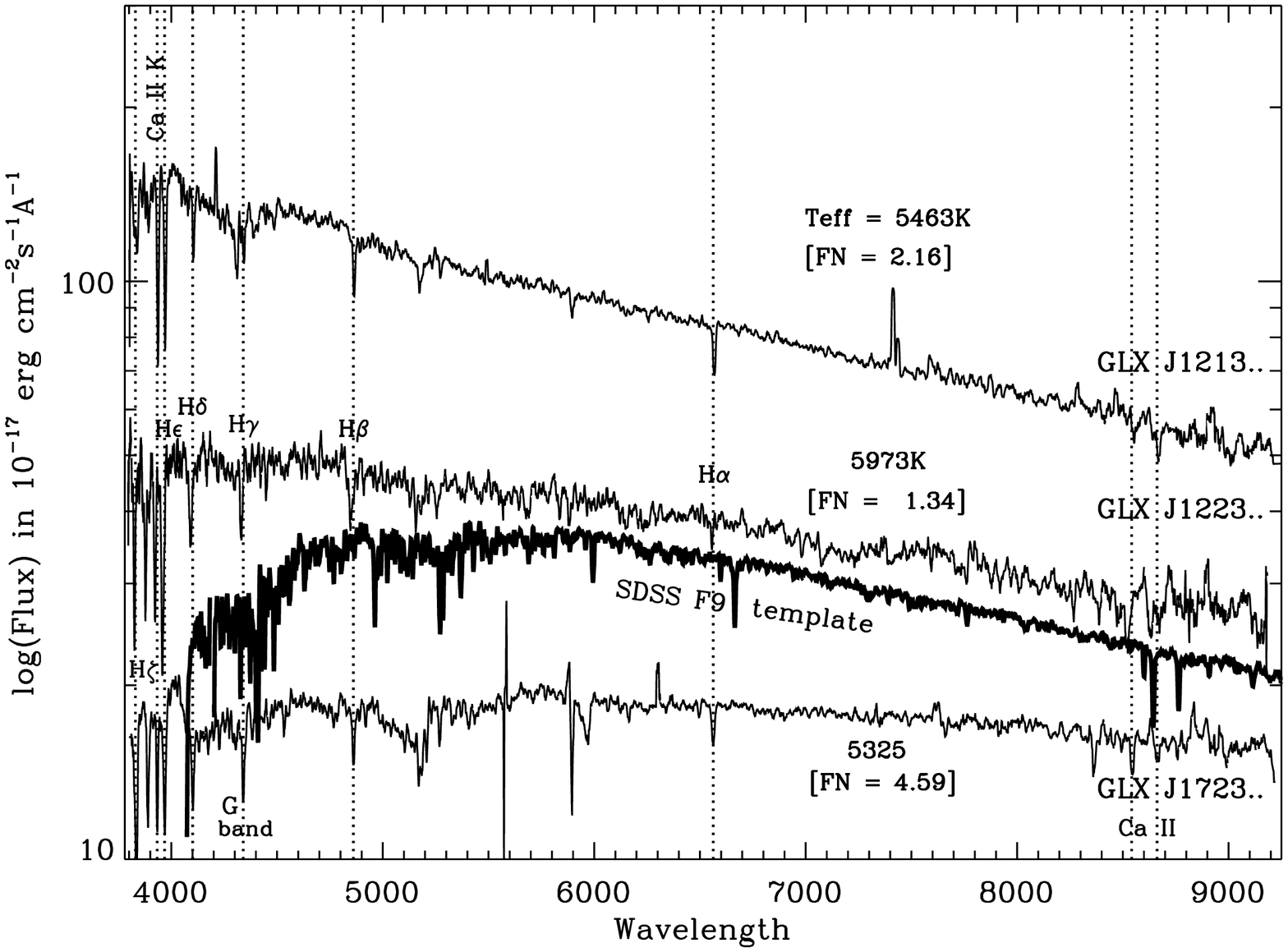}} 
\caption{Light solid: SDSS spectra of a subset of UVe stars shown in 
Fig.\,\ref{fnteff}. Thus, annotated T$_{\rm eff}$s are those from the 
SDSS/SSPP/DR7.  Fluxes are smoothed by a running 8-point boxcar function. 
For convenience of display the middle and top spectra have been rescaled. 
Also, the spectrum in bold for clarity exhibits a 
SDSS F9V template, shifted by +100\AA\.~
Dotted lines mark the positions of the Balmer and Ca\,II lines
%He\,I, 
and G band. Notice the fluxes at the short wavelength end of
the lower spectrum for the F9 template and mildly UVe object for GLX
J172312.5+593928 taper, whereas continuum fluxes continue to rise 
the topmost two UVe spectra.  
%%(The Mg\,I/MgH feature at 5100-5200\,\AA~ 
%is not used for spectral typing by the SDSS pipeline.)
}
\label{3sdss}
\end{figure*}

%With the point clearly established that the UV spectra of the
%photometric UVe objects also show unexpected rising slope 
%towards shorter wavelengths, 
To complement the GALEX UV spectra, we checked the SDSS archives with
the on-line Spectroscopic Query Form to examine the appearance 
of optical spectra.
We have used this tool to compare a number of spectra on either
side of the UVe/UV-normal demarcation (dot-dashed) line in 
Figure\,\ref{fnteff}. For brevity we display in Fig.\,\ref{3sdss} 
the first and second spectra in Fig.\,\ref{2pnl}, i.e., the near-UV normal
spectrum of GLX J172312.5+593928 as well as examples of spectra
of two UVe stars, GLX J122321.1+155205 and GLX J121309.5+504539. 
The SDSS F9 template spectrum is also shown for reference.\footnote{SDSS
spectral cross-correlation templates can be found at
http://www.sdss.org/dr7/algorithms/spectemplates/.}
% (GLX J122321.1+555205 and GLX J121309.5+504539, respectively). 
%The top and middle spectra
%in particular show more negative UV continuum slope
%than even a type A7 spectrum shows (see Fig.\,\ref{bertn}).
These examples summarize most though not all of the differences
we find in these objects. 
In the great majority of cases,
and allowing for equal effective temperatures, we find that
blue fluxes are greater in the UVe stars. 
% than examples occupied by the
%normal population of stars to the upper right of Fig.\,\ref{fnteff}. 
%In the spectra of the UVe stars in Fig.\,8
%the continuum fluxes continue to rise in the available blue segment
%at 3800-4500\,\AA.  

% FIGGURE 8 ; 3 SDSS spectra - 3sdss

%Changes above: still need to change in plot 6620K --> 5325K, 5463K --> 5574K
% changed "middle spectrum (to GLX1213) to "middle spectrum"

%In addition, 
Our SDSS spectra show similarities with those given by 
Yanny et al. (2009). In particular, their spectra of ``F/G" and "G giant" 
stars (their Figs. 9) are similar to our reference standard. 
However, our UVe star spectra exhibit 
%For example, the continua of their blue straggler A, DA and sdB stars 
%%(their Figs.\,4 \& 6) have some gross similarities, but the appearance of a 
somewhat a weakened Ca\,II K and IRT lines and G band with respect to the
F9 template. 
%% and occasionally the presence of a He\,I 3889\,\AA~ line 
%%in our spectra of UVe stars 
%%This suggests either a continuum contribution from a hot source or a higher 
%%atmospheric temperature minimum.
%Findeisen, Hillenbrand, \& Soderblom (2011) report a correlation between 
%GALEX FUV flux and K line emission. (For completeness, neither 
%the SDSS spectra of our exemplars nor other UVe stars in the GALEX-SDSS 
%exhibit K line reversals.)
%%Such reversals would be expected in spectra of only the youngest solar-type 
%%stars.  
% which could arise from contributions from a chromosphere or a % hotter 
% secondary star.  
According to Fig.\,\ref{3sdss}, the line weakening extends to the Ca\,II 
triplet in the IR as well.
This indicates that the line weakening is not necessarily entirely
limited to short wavelengths. Indeed, we would expect that the cores of all
these Ca\,II lines would be sensitive to the raising of the atmospheric 
temperature minimum associated with a strong chromosphere. 
%The red end of the spectrum supports the view that any contamination is
%limited to short wavelengths.  
%Also, whereas SDSS spectra of A type stars exhibit Paschen lines in the 
%wavelength region 8800-9000\,\AA,~ these features are not visible in the 
%UVe spectra. 
%Instead, the spectra of the UVe stars exhibit 
%two of the stronger lines of the Ca\,II IR triplet (IRT), characteristic
%of a G but not an A type star. 
%Note that the IRT is formed at the temperature
%minimum of the atmosphere, the same as the G band. This fact suggests after all
%that a well developed chromosphere does not form this pattern of features.
All told, there is good evidence for a mild flux enhancement in the 
blue region of these optical spectra, either from a secondary star or a
strong chromosphere. 
%If a binary explanation explains the UV excesses, 
%the inferred spectral types of the secondary companions must be
%type A or B stars, i.e., similar to those inferred from the slope of the 
%UV continuum in the GALEX spectra. 
%%We shall refer to 
%% the UVe objects as ``giants" or ``binary dwarfs" as 
%%the UVe objects as ``UVe giants" or ``UVe dwarfs" as 
%%appropriate in the following.

\subsection{Adding a 2MASS color to the mix}
\label{uvoir}

\begin{figure}[ht!]
\centerline{
\includegraphics[scale=.37,angle=90]{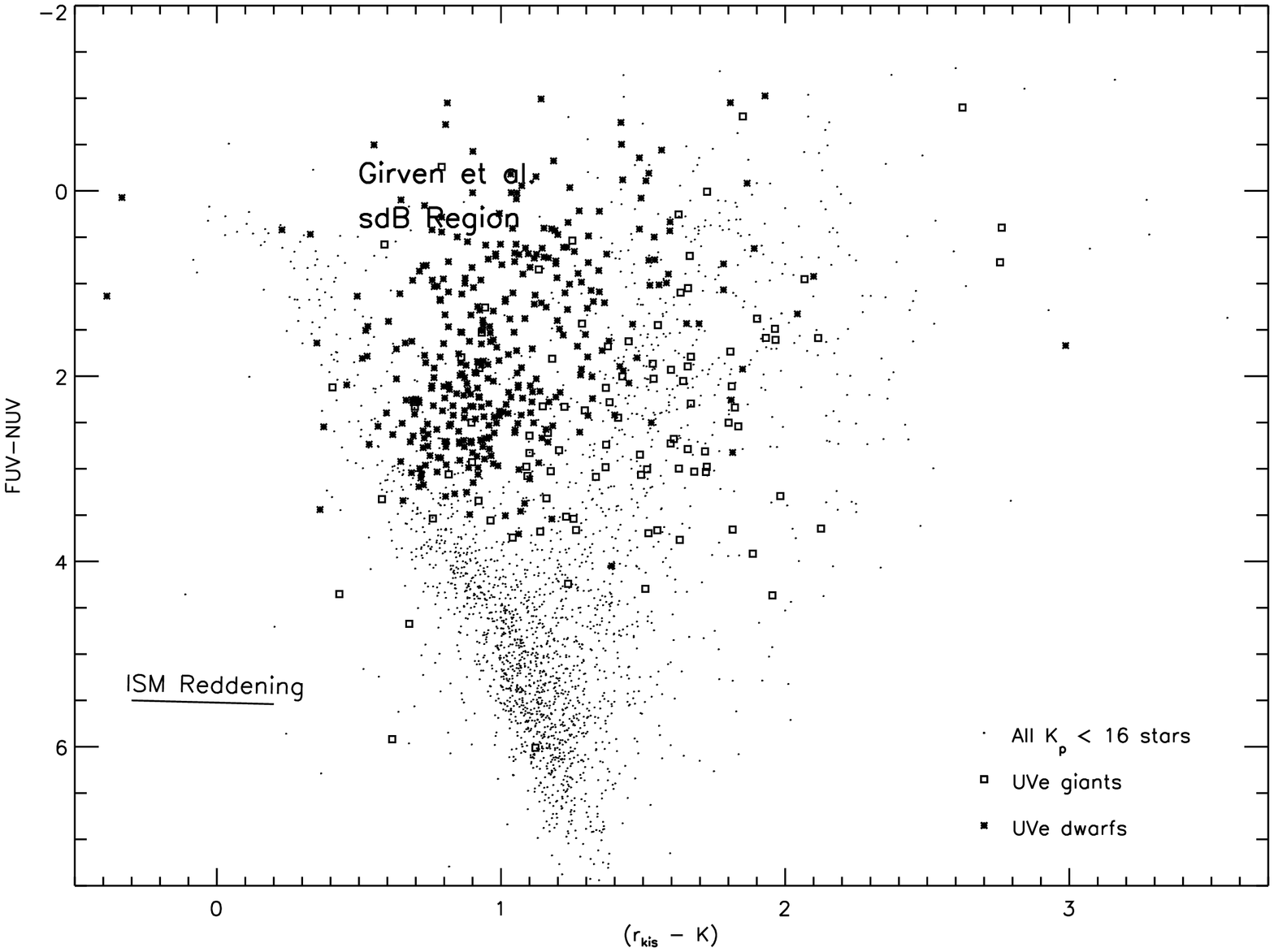}} 
\caption{ UV/optical/IR (r$_{kis}$ -K), (FUV-NUV) diagram showing UVe stars.  
All magnitudes 
%are plotted according to their catalog values -- all but
but r$_{kis}$ are in the AB system.
The diagonal running from upper left to lower right represents the main 
sequence. The high density of small points beginning at (r$_{kis}$ -K) 
$\approx$ 0.8 consists of F-K type stars in the upper right of 
Figs. \ref{fnteff} and \ref{fngi} while the scatter of the crosses and open 
squares in the upper right of this figure represent UV-normal and UVe giants. 
The sdB star-occupied region (after Girven et al.) is shown. }
%(2012b), 
\label{rkfn}
\end{figure}

 Another trait of the UVe population is their displacement from 
the ``UV normal" stars in a 2-color UV/optical/IR diagram. 
In Figure \ref{rkfn} we replot the population from Fig.\,\ref{fngi} 
in an (r$_{kis}$-K), (FUV-NUV) diagram. Here the K magnitude refers to 
the 2MASS filter centered at 2.2\,$\mu$m.  
We see that even for the stars with small UV colors
% above line added on 7/1/13
the distribution of UVe stars does not match the normal star distribution. 
The principal effect is that the (FUV-NUV) colors of the UVe stars 
are bluened by 4 magnitudes or so. Also,
the $(r_{kis} - K)$ optical/IR colors of the UVe
giants are displaced to the red by at least 0.5 magnitudes. 
This shift appears to mirror the displacement of the UV-normal late-type
giants in this diagram by about +1 magnitude. It is likely an aspect
of the intrinsic redness of near-IR colors of late giants. 
%as well as to these stars' larger foreground ISM reddenings.

% FIGGURE 9 -

Because of the blueing of their (FUV-NUV) colors, only a small fraction of
the UVe stars from Fig.\,\ref{fngi} fall on the diagonal upper main sequence 
band in Fig.\,\ref{rkfn}. This may be due in part to additional reddening of 
the (r$_{kis}$-K) colors of the UVe stars, as with their $(g-i)$ colors.
%Even so, and importantly,
%the (FUV-NUV) shift is not quite as large as the computations of 
%Girven et al. (2012b) show for the sdB stars on this diagram. 
%We have annotated the sdB region in our figure to highlight the only 
%small overlap of the two populations. 

%However, the reverse is not quite true.
%5A number of ``normal," particularly early-type giants from 
%Fig.\,\ref{fngi} are reddened with respect to the main sequence 
%stars and overlap the UVe ``binary" domain in  Fig.\,\ref{rkfn}. 
%It turns out the we will be able to identify these below as a separate
%group of stars.
%% mainly, single horizontal branch AF stars.

\subsection{UV-red clump: segregating AF III and UVe stars}

  Smith \& Shiao (2011) first noticed a UV-red ``clump" island in the 
$(g-i)$, $(NUV-g)$ color diagram. This clump
feature is composed of stars with (FUV-NUV) $<$ 2.5 and $(g-i)$ values 
consistent with A and early F stars. But why should AF stars be surrounded 
within a sea of UV-blue stars, including stars with red optical colors and
therefore low effective temperatures?

\begin{figure}[ht!]
\centerline{
\includegraphics[scale=.35,angle=90]{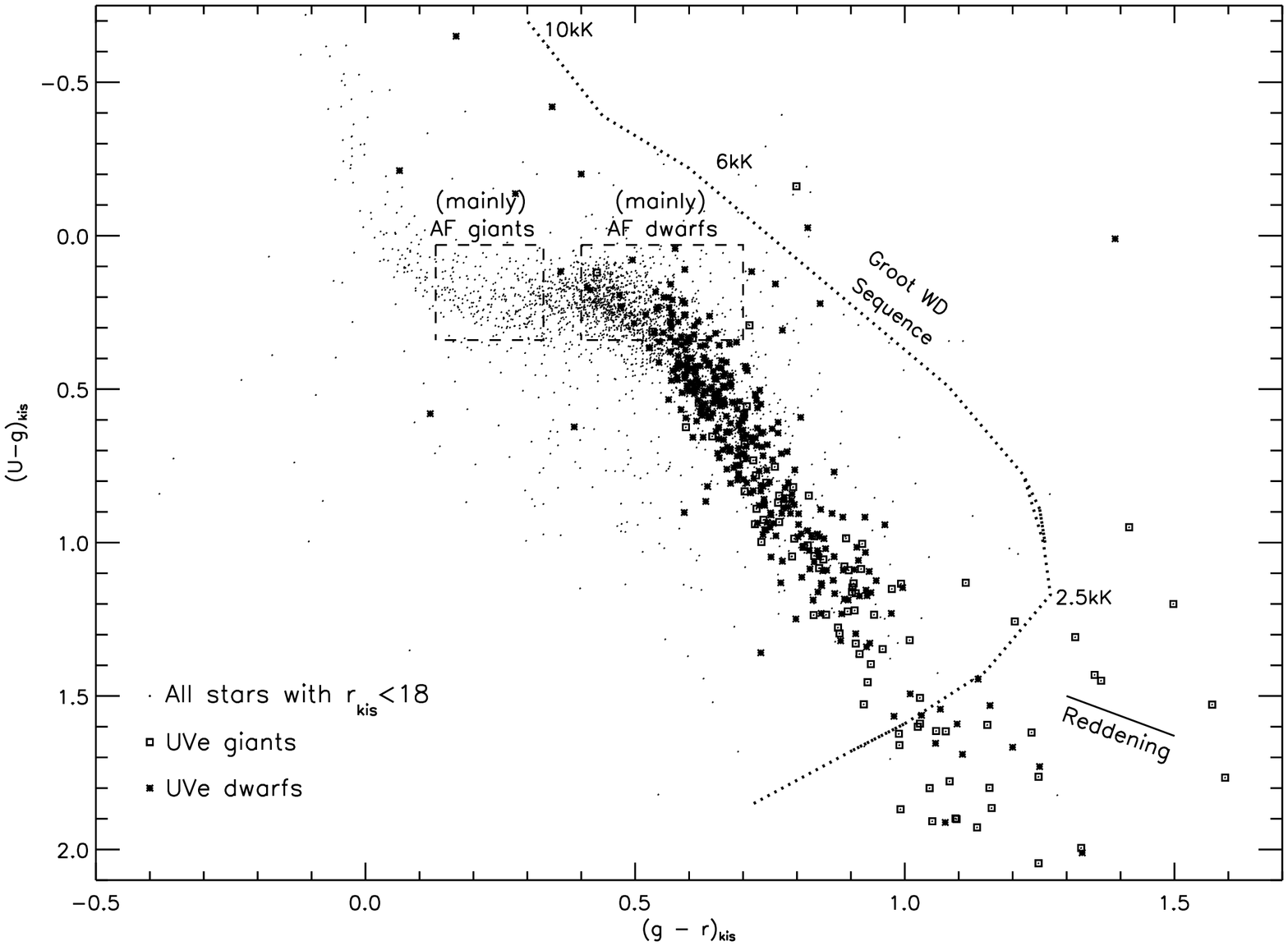}} 
\caption{The SDSS $(g-r)$$_{kis}$,  $(U-g)$$_{kis}$ diagram
%, similar to the Johnson $(B-V)$, $(U-B)$ diagram, 
for all stars with K$_p$
$<$18 for stars in the Kepler field. Special symbols denote the 
UVe giants and dwarfs (as flagged in Fig.\,\ref{fngi}) 
and show again that the UVe sequence 
follows an extension of the normal dwarf and giant star sequence for
intermediate and low mass stars. The regions defined by dashed boxes
consist primarily of AF dwarfs and giants. 
The Groot color sequence for white dwarfs is plotted.
%(Vega system) for white dwarfs (Groot et al. 2009) is  depicted by dotted lines.
% does not match the observed positions of these stars.
}
\label{grug}
\end{figure}

We addressed this question by first looking at the behavior of stars in 
the Kepler FOV in a SDSS-filter version of the Johnson $(U-B)$, $(B-V)$ 
diagram - see the $(g-r)$, $(U-g)$ plot (Figure\,\ref{grug}).
%observed in the SDSS/KIS system.
Straizys et al. (1998) have computed synthetic SDSS color indices for 
the main population sequences.  As with other photometric systems, 
%employing near-UV and visible band filters, 
their work demonstrates the usefulness of the
Balmer jump in separating A-F main sequence from higher luminosity stars.
% (see their Fig.\,2). 
In particular, similar to what happens in the 
$(B-V)$, $(U-B)$ color diagram, the effect of the Balmer jump is to make a 
marked kink in the $(g-r)$, $(U-g)$ diagram, such that giants and especially 
dwarfs have nearly the same $(U-g)$ for a range of $(g-r)$ and $(g-i)$ values.
Figure\,\ref{grug} shows these colors, again for KIS survey 
stars with K$_p$ $<$16; UVe giants and dwarfs from Fig.\,\ref{fngi} are
denoted by separate symbols. We notice first that
the UVe population coincides with the sequence of cool
UV-normal stars. Second, the UVe stars, whether giant or dwarfs, 
are much redder in $(g-r)$ than UV-normal AF stars. 
We divided the region occupied by A and early F stars into
Fig.\,\ref{grug} into two boxes, indicated by dashed lines.
According to Straizys et al. these 
subregions are expected, from left to right, to be occupied mainly by giants, 
a mixed population, and giants, respectively. In fact, the fractions of 
giants in these boxes turn out to be 70\%, 43\%, and 30\%, respectively, 
%dwarfs in these boxes turn out to be 30\%, 57\%, and 70\%, respectively, 
according to our K$_p$ and KIC-derived \logg~ criteria.
% giant subregions are identified in the diagram by dashed boxes. 

% FIGGURE 10 ; g-r, U-g

\begin{figure}[ht!]
\centerline{
\includegraphics[scale=.40,angle=90]{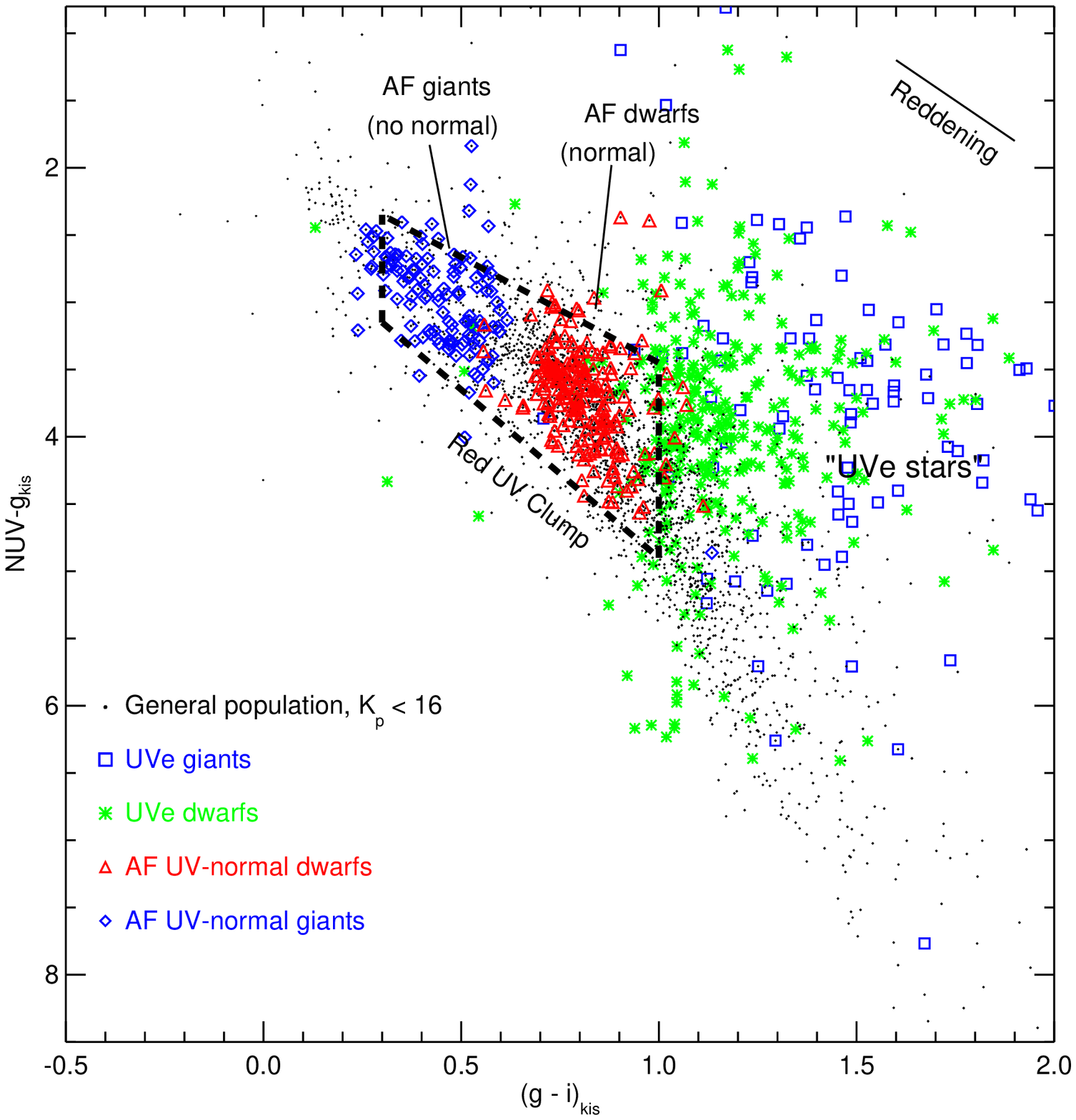}} 
\caption{ The $(g-i)$$_{kis}$, $(NUV-g)$$_{kis}$ diagram for stars within
the Kepler FOV with K$_{p}$ $<$18 magnitude. The prospective UVe dwarf 
and giant stars from the previous diagram are now color coded. The stars 
in the left and right boxes region in the previous figure 
are referred to as AF giants and AF dwarfs, respectively.
The trapezoid in Fig.,\ref{fndist}, indicating the position of the ``UV-red 
clump" is again shown in this figure, but the $g$ magnitude and $(g-i)$ 
color are transformed to the Vega magnitude system. For purposes of 
clarity we have plotted every second and fifth member of the UV-normal 
AF giant and dwarf groups, respectively.
}
\label{ging2}
\end{figure}

All stars within the giant-box and dwarf-box regions of Fig.\,\ref{grug} 
were flagged and are indicated in our $(NUV-g)$, $(g-i)$ diagram 
(Figure\,\ref{ging2}) as blue diamonds and red triangles, respectively, 
whereas stars formerly marked UVe dwarfs and giants 
appear there as green asterisks or blue diamonds, respectively. 
Note first that the UVe giants are significantly redder the 
UVe dwarfs,  similar to the reddening noted in Fig.\,\ref{fngi}. 
We see also that some 84\% of the stars occupying the UV-red clump 
region (the trapezoid in Fig.\,\ref{ging2}) are AF stars. 
The other 16\% in the clump region have SDSS \teff\, values, where they 
have been determined, consistent with late B or mid to late F spectral types. 
The fractions of the population spilling within and without 
the two trapezoid areas are consistent with the overlap of
a wing from one (FUV-NUV) distribution shown in Fig\,\ref{fndist} with the
peak of the other. 
The temperature range represented in this region is \teff~ 
$\approx$ 6.5-9\,kK. Some 70\% of the total of 
1785 stars in the UV-red clump zone are giants. 
This is virtually the same value, 71\%, that Smith \& Shiao (2011) found
for UV-red clump stars observed in their independent sample. The numbers 
of dwarfs and giants in the middle region of the trapezoid (left uncolored 
in Fig.\,\ref{ging2}) are about equal.
% common GALEX-SDSS sky field.
  
% FIGGURE 11 ; NUV-g, g-i

  Does the UVe population avoid the UV-red clump? 
The answer is indeed yes.
Only 1.4\% of the UV-red clump stars were flagged
as UVe in Fig.\,\ref{fngi}. The colors 
%of the ``binary" population, dwarfs, and even moreso giants, are 
of the UVe population, dwarfs and even more so giants, are 
displaced to the red in $(g-i)$ of the trapezoid region. 
%The giants are displaced relative to the dwarfs in the 
%clump region, and this is a restatement of their redder $(r-K)$ and
%$(g-r)$ and IR colors in Figs.\,\ref{rkfn} and \ref{grug}.

 In sum, the UV-red clump region is mainly occupied 
by UV-normal giant and dwarf AF stars.
These are the stars in the upper left (calibrated UV-normal relation)
of Fig.\,\ref{fngi}.  These populations
%, main sequence turn-off and giant AF stars, in fact 
have blue (FUV-NUV) and $(g-i)$ colors that are consistent 
with the temperatures expected for normal single stars.
%%Most importantly, they are putatively {\it single} 
%%stars - not members of our UVe ``binary" population. 
%%Since their colors are not contaminated by the ``binary" group's UV excess, 
%in terms of the color scheme in Fig.\,\ref{ging1}, their GALEX UV colors 
%are ``red" of our initial (FUV-NUV) color demarcation. 
%Put another way, the AF horizontal branch stars would
%Put another way, the AF horizontal branch stars would
%have red (FUV-NUV) colors too, but for the population of UVe
%(mainly F) stars that have been added to the same (FUV-NUV) color 
%range in Fig.\,\ref{fndist}. It is the ``binary" stars that exhibit
%actually UV color excesses. They are numerous enough to cause
%the mean color of the A and F stars in 
%Fig.\,\ref{ging1} to shift from blue to red.
In contrast, a comparatively large fraction of late-type stars are
UVe stars. When these are added to the
temperature class bins in Fig.\,\ref{fndist}, their influence is 
enough to move the mean (FUV-NUV) values markedly, causing the 
4\,kK and 5\,KK symbols to march back to the blue in this diagram.

\section{Discussion: Explanations for the UVe population}
\label{discs}

\subsection{Confusion with unresolved extragalactic objects?}

 To address this question, we recap by first stating that we used the 
GALEX and SDSS galaxy-star separation flags to assemble our populations. 
Both the GALEX and 
SDSS pipelines assess whether the image of an object is point-like. 
% i.e., whether it is a star or QSO as opposed to a partially resolved galaxy. 
This flag is robust for the SDSS survey down to $r$ $\approx$ 
21.5 mag. (e.g., Lupton et al. 2002). 
%Scranton et al. 2002). 
The limit for
GALEX is $\sim$3$\frac{1}{2}$ magnitudes shallower, but still deeper
than the 16.0 magnitude limit we imposed.  A similar
flag was used for the KIC, though it is taken from a variety
of earlier photographic catalogs. Based on these flags
and the review of spectral features the pipeline judges to be stellar,
there is little chance that many of our stars 
could have been confused with even small resolved galaxies. 
%Their redder colors and brightnesses tend to exclude them as well from QSOs. 

As a second consideration,
Bianchi et al. (2007, 2009, 2011a) have examined the distribution of objects 
in the GALEX-SDSS sky with respect to the GALEX UV color and the SDSS $(g-r)$
and $(g-i)$ colors. They found that both hot and cool stars and QSOs 
with redshifts $z$ $\sim$ 1 are well separated in these diagrams. For example,
In the $(g-i),$ $(NUV-g)$ diagram (Fig.\,4 of Bianchi et al. 2011) QSOs
inhabit the upper left region of the plot, i.e., ($(g-i)_{kis}$ $\gtrsim$ 0.7,
$(NUV-g_{kis}$) $\ltsim$ 2. This just abuts 
the region inhabited by UVe stars, according to Fig.\,\ref{ging2}. This just
The QSOs principally fall in a box indicated by
dotted lines in Fig\,\ref{fngi}. One can see that this population
falls in between the UVe and UV-normal stellar populations in this diagram. 
Although it is possible that it would include a 
handful of objects in the UVe zone, we consider that QSO colors would not 
include most of the much larger sky area where our UVe objects fall. 
From this consideration we estimate the contribution to be $<$ 1\%
of our detected UVe objects. The upper limit here comes from the concentration
of the extragalactic objects in a smaller color region and the fact that
these objects are generally much fainter than the UVe stars.
%We will discover this same mismatch in our discussion of hot WDs.

Third, Bianchi et al. (2007) compiled a census of objects for the GALEX
MIS and SDSS surveys as a function of the $r$ magnitude (close to the
Kepler K$_p$). Scaling their numbers of objects found to $r$=16 in the 
overlapping region, 86 deg$^2$, to the overlapping Kepler/GALEX field,
one predicts a total of about 587 unresolved galaxies and/or QSOs. 
This number drops to about 460 if the difference in ISM reddenings between
the two sky regions ($\Delta$A$_{FUV}$ = 0.5 mags) is taken into account.
% (actually it is 0.6 mags)
The figure 460 falls short of the estimated $\approx$1012 
(2$\times$506; 506 from $\S$\ref{glxcol}) UVe objects
%order of magnitude short of 786 
% 451 anom-UV objs in KIS1 field of 50 deg^2
in the full KIC field. So while the number of objects almost coincides,
the shortfall is still significant.  Because the Bianchi et al. numbers
vary across the sky, and we have taken an average, we once again cannot 
completely rule out a minor contribution. 
% <EBV>=.173,.185 (for KIS1, KIS_total) vs. <EBV>=.041 for GALEX-SDSS
%our statistics (451 UVe objects). 
%A related estimate to the problem, more relevant for UV objects can be
%made by assume
%limiting values $r$ = 16 and (NUV - g) = 4 of putative objects within the
%5 arcsec search radius of our objects and to ask what numbers of UV objects
%might be detected in the NUV band of GALEX. Scaling magnitudes and search
% radii values from Figs. 3 and 4 of GALEX suggests that as many as 1\% of
% obects might be detected.....

Finally, we will find that the fraction of UVe objects, 17-22\%, in the Kepler 
FOV and GALEX-SDSS, magnitude-limited skies are, within errors, the same 
as the much more carefully vetted (and generally apparently brighter) 
members of Eclipsing Binaries and Kepler Objects of Interest, also 
discussed in $\S$\ref{glxcol}. 
All members of these two groups are unquestionably stars.

For these reasons we reject the notion that a significant fraction 
($\gtrsim$1\%) of the UVe objects is extragalactic. 
%We believe the objects are 
%overwhelmingly stars, as expected also from the restricted brightness range
%of our sample, and our discussion below will proceed from this result.

\subsection{``False positive" stars along the line of sight to UVe objects}

  If the UVe of this population are in fact caused by
foreground/background hot stars along the line of sight to a sky area 
subtended by the ground-based Kepler and KIS-extracted aperture of a cool 
star, this might explain the apparent contradiction in UV and optical band
photometric colors.  Several astronomers working directly with the 
Kepler Project have considered whether such chance alignments, 
so-called ``false positives," could be responsible for exoplanetary 
identifications among the Project's list of ``Kepler Objects of Interest" 
(KOIs). A recent simulation of the false positive occurrences for the 
KOI list by Fressin et al. (2013) predicted a rate of 9.4\%.
So, it is possible in principle that false positives could play a role in the
positions of our UVe populations in the GALEX-SDSS two-color diagrams. However,
this is actually not likely because false positives for the KOIs invariably 
refer to transits or interference by faint {\it cooler} stars, either red 
giants or lower main sequence stars. Since the space densities of hot stars
are much lower, we do not expect hot background stars to contribute much 
to our UVe population. Likewise, hot WDs and subdwarfs have lower 
space densities (as well as smaller sampled volumes).
Hence we rule out false positives too. 
% as a significant constituent of the UVe star population.

\subsection{Could the UVe stars be WD or hot subdwarf binaries?}

  Because of their high temperatures the primary candidates for UV excess 
objects in recent surveys have invariably turned out to be WDs and sdOB 
stars, whether as single stars or primaries in binary systems.  
However, several published studies suggest that although WDs are
an important constituent of the UV-excess population, they are not 
the right match to the UVe stars from our color diagrams.
We take the WD group first and follow it by another
candidate group, the sdB stars.

\subsubsection{White dwarfs and single subdwarfs as candidates}

After defining the loci of the main stellar populations in 
GALEX-SDSS color-color diagrams, Bianchi et al. (2007) showed 
that hot and warm ($\gtrsim$18\,kK) WDs can be separated from other
stellar populations according to their UV colors. 
One then asks could WD single or binary systems be the cause of the 
UV excesses of the UVe stars? The predicted WD locus in their 
$(g-i)$,  (FUV-NUV) plot suggests that a simple extrapolation of the
colors of the WD sequence might overlap the UVe class. 

Yet, closer examination of extant WD star lists establishes
that this explanation is not promising. 
%For example, 
An unpublished list by Ostensen (2013) of 24 warm to cool WDs 
classified from stellar spectra shows that the stars are too faint 
(K$_p$ $>$ 18) and/or fall in the wrong section of the (FUV-NUV), $(g-i)$ 
diagram to be considered as UVe members. Only three of the Ostensen 
objects fall within the upper left edge of the observed distribution, 
and all of them are fainter than the K$_p$ = 16 magnitude limit we
considered for our investigations. Cooler WDs do not 
continue this trend because their red and near-IR colors 
% (e.g., $(g-i)$)
do not continue to increase. 
%This is because of the mutilation of the 
%red fluxes due to collisionally-induced absorption from molecular $H_2$ 
%in the stars' atmospheres. 
In addition, we find that the computed
WD sequences in the $(g-r)$, $(U-r)$ diagram
%, similar to their behavior in the Johnson $(B-V)$, $(U-B)$ diagram, 
follow a redder sequence in
$(g-r)$ than do the main sequence and fainter stars (e.g. Verbeek 
et al. 2012b). A similar mismatch is depicted by the WD locus
computed by Groot et al. (2009) in Fig.\,\ref{grug} (dotted line). 
For cool WDs the computed locus 
lies well to the red of most of the observed colors of our sample.

The Girven et al. (2012a) catalog of single DA white dwarfs verifies 
this conclusion with a much larger sample.  The subset of this catalog
population with GALEX and SDSS colors totals 12\,149 stars. 
Only 0.18\% of this population 
%(and a considerably smaller percentage of 
%the K$_{p}$ $<$ 16.0 population we are studying) 
fill the upper left edge of the UVe region in our (FUV-NUV), $(g-i)_{kis}$
diagram ($(g-i)_{kis}$ $\le$ 0.7) 
%and also satisfy the brightness constraint
%K$_{p}$ $<$ 16.0 used in our display of UVe stars in Fig.\,\ref{fngi}. 
%On the other hand, the percentage would increase to 
%2.3\% if a criterion K$_{p}$ $<$ 18.0 were to be used - still a 
%small value compared to the UVe population. 
Nearly all DAs from Girven et al. fall in the upper left 
% (``UV normal star") 
region of Fig.\,\ref{fngi}, where the Bianchi et al. 
(2007, 2009) diagram predicts they should be. 
A handful of cooler stars (\teff = 6.5-9\,kK) falls to the upper left 
(defined again as $(g-i)_{kis}$ $\le$ 0.7), but none of these spill 
into the UVe zone. 
%In fact they are found well to the right 
% ((FUV - NUV) $>$ 3), in the UV-normal population zone.  
In a similar $(r-K)$, $(NUV-r)$ diagram, Girven et al. show that 
the DA stars have $(NUV-r)$ colors that are 1 to 5 magnitudes bluer than 
{\it all} objects in our KIS sample, including our UVe population. 
%Similarly large $(U-g)$
%super-excesses can be found in the UVEX catalog relative to our UVe
%population (Verbeek 2012b). 

This discussion assumes that catalog objects are either single stars 
or binary systems in which the late-type secondary companions 
%dominate or 
contribute substantially to the optical/near-IR 
%(SDSS) 
colors.  Indeed, the fact that we find that the DA stars are restricted 
to a blue region, $(g-i)_{kis}$ $\le$ 0.7, in Fig.  \ref{fngi} 
would be consistent only with secondaries having 
early K0 spectral types or earlier.

 Other instructive resources are the Rebassas-Mansergas et al.
(2010, 2012)  catalogs of white dwarf- main sequence (WDMS) systems. 
These binaries consist of visible wavelength primaries and dM secondaries. 
%In fact, the dM secondaries are visible only in wavelengths beyond 7000\,\AA.~ 
These authors' $(g-i)$, (FUV-NUV) figures of WDMS systems in the 
SDSS-GALEX sky
show almost no overlap with our UVe zone. 
Rather, the UV excesses are significantly larger than our group, 
the only stellar class in our GALEX-SDSS survey for which this is true. 
%Thus, the (FUV-NUV) color is typically 2 magnitudes 
%bluer for the Rebassa-Mansergas et al. sample than our UVe members.
Moreover, their WDMS have magnitudes of K$_p$ = 19-20.

In addition, the space density of {\it all} WDs in the solar 
vicinity is estimated to be 3.2 $\times$10$^{-3}$ M$_{\odot}$\,pc$^{-3}$
(Holberg et al. 2008), or only about 10\% of the local density of lower 
main sequence stars (Gilmore \& Zeilik 2000). 
%Furthermore, of all WDs
%only $\sim$30\% are expected to be in binaries or multiple systems 
%(Holberg et al. 2013).

In sum, neither single white dwarfs nor WDs in binaries with late-type 
secondaries are bright and numerous enough to contribute much to the
UVe population.

\subsubsection{The subdwarf B stars }

  Another interesting population to compare to our UVe population is
the class of subdwarf B (sdB) stars. These core He-burning stars are 
found on the extreme blue end of the horizontal branch (eHB)
 and are thought to reside there for $\sim$160\,Myrs. Thus, they form a 
rarer, though more luminous, population than the hot DA or DB stars. 
%They overwhelm WDs in UV surveys and are thought to be 
%responsible for the UV excess in elliptical galaxies (e.g., O'Connell 1999). 

%According to Maxted et al. (2001), 
%their radial velocity and photometric colors indicate that 
%40-70\% of sdB are in binary systems.
Evolutionary models suggest a few possible routes to the formation of sdB 
stars, usually in close binaries 
%, for example, in common envelope mass transfer or WD+WD merger systems 
(e. g., Han et al. 2003).  However, it is unclear from observations 
that the observed frequency (40-70\%) of short-period sdB binaries 
is as high as the  models predict (Barlow et al.  2012). 
%However, it is possible that a small percentage may 
%evolve to this state as single post AGB stars that have been spun up 
%by angular momentum transfer by a companion star, exoplanet, or 
%internal rotation of the core (Politano et al. 2008). Eventually most 
%will evolve to become a class of WD binaries. 
Some studies have reported that a substantial fraction of sdB 
binaries have F-K main sequence secondaries. For example, decompositions 
of binary sdB spectra suggest typical cross-overs 
for the fluxes of the two component stars in the blue-yellow wavelength
range (Aznar Cuadrado \& Jeffery 2001, N\'emeth, Kawka, \& Vennes 2012). 
%Girven et al. 2012b).  
Even though they emit primarily in the UV, the
sdB components of such systems are so bright that usually they are the 
visible band primaries. 
%finds that a substantial fraction of sdB stars, like the WDMS systems 
%of Rebassas et al. (2010, 2012) have F-K main sequence secondaries. 
%Added to the mix are a relatively few systems with type A and M dwarf 
%secondaries.  This distribution is similar to the main sequence sampling 
%of our ``binary" population. 
%Consequently, rather different spectral types result from assessments of
%spectral features in the near-UV/blue or yellow/red spectral regions
%(Girven 2012b,  N\'emeth, Kawka, \& Vennes 2012). Notably, this is not
%true of the far-UV versus the visible wavelength band.

%  The discovery of sdB-late main sequence binaries has been bolstered
%by the UVEX survey and in particular the follow-up UV/visible/IR survey 
%of Girven et al. (2012b). 
%Flux decomposition of visible bandspectra of a number of
%photometrically discovered sdB binary candidates by N\'emeth, Kawka, 
%\& Vennes (2012) as well as photometric color decomposition by Girven
%et al. shows that UV/visible/IR colors confirm that a number of
%sdB systems have visible band companions and give mutually
%contradictory information on spectral types.

  There are problems matching the detailed color anomalies of sdB binaries 
beyond those we have reported. The simulated sdB-dwarf binary 
zone in model binary SEDs by Girven et al. (2012b) is shown in
our Fig.\,\ref{rkfn}. One can see here that the distribution of 
our UVe population stars falls to the lower right of where these models
indicate they would be.  One can imagine a better agreement if the
secondaries of sdB systems were late-type giants because this would redden
the colors, moving the model locus to the right in the figure. 
However, we have seen that the fraction of giants among UVe stars
is at most 22\%. From Girven et al's models best
agreement with our colorimetric results is met when the UV 
excesses are smaller and hence the radii of the contaminating hot 
secondaries smaller. According to their work the best match in 
color-color diagrams for hot degenerate-main sequence binaries occurs 
for small radii, coresponding to 
the requirement \logg\,$\gtrsim$\,7. But such high \logg~ values obtain 
only for WDs, not sdBs, and we all but ruled these out 
because of their faintness and low space densities.

%Despite their similarities to our UVe stars in terms of 
%the contradictory information implied from the UV and visible bands, 
The space density of sdB objects is only
$\sim$10$^{-6}$ M$_{\odot}$\,pc$^{-3}$ (Nelemans 2013). 
% or $\sim$ 0.1\% of the mass density of Galactic disk main sequence stars.
Then the sdB population fails by two orders of magnitude to explain 
the UVe stars.

\subsection{Other explanations }

\subsubsection{Color fitting with model binaries}

 Considering the presence of excess blue flux 
(Fig.\,\ref{3sdss}), our focus on the cause of the 
UV color excesses should be directed towards  populations that are almost
comparable to the full UV-normal population itself. We have mentioned two 
such explanations. The first is that the chromospheres are strong
(e.g., stellar youth). The 
second is subgiant-main sequence binaries with the components
having only slightly unequal masses (e.g., blue stragglers). 
Such systems would have hot secondaries on the main sequence that dominate 
the UV and evolved primaries, possibly aged main sequence stars,
subgiants or just evolved giants.
The systems favor primaries being cool giants, for which 
temperature differences are greatest. However, the 
contribution from the hot dwarf secondaries would then be less because 
of their smaller radii.  We consider here whether a viable parameter 
space exists for the nondegenerate binary scenario. 
%to cause the color shifts observed.
%However, all is not well with either of these explanations.

%The mass ratios of such systems would be near unity.
%On balance, some support for this picture can be found from the Bayesian
%approach the fitting by Janes et al. (2013)  member stars of the 
%young cluster NGC\,6866, also in the Kepler field, to a common isochrone. 
%These authors found a number of G main sequence stars that are likely in fact
%to consist of binaries consisting of G subgiants with secondaries having
%nearly the same mass.

  To address this question we tested whether fluxes of less evolved, hot
secondaries dominate the GALEX UV colors of binary systems. 
We did this by folding the GALEX and SDSS filter transmission curves of 
each star, in the matter described by Bianchi (2009), and evaluating 
the summed fluxes in each filter. 
%The computer program to do this was described by Bianchi (2009). 
%convolved through of their combined fluxes in 
%the bandpasses of the GALEX and SDSS filters.
% We return to the question posed in $\S$\ref{glxcol} of whether the UVe
%stars can be optical-UV binaries in which the primary is a partially
%evolved F-K main sequence to subgiant star and the secondary is a hotter 
%star not yet evolved much from its unevolved position as an A or B main 
%sequence star. Such binary configuration are not expected to occur with
%great frequency, but we need to know first whether the colors can be
%altered in such a way that the UV and visible wavelength colors give
%photometric spectral types that are discordant enough to show the UVe
%distribution particularly in Fig.\,\ref{fngi}. 
%In particular, we expanded the model grids described by Bianchi (2009) and
%computed colors from the known GALEX and SDSS filter transmission curves  
%model atmospheres
To carry out these tests we took atmospheric flux models for standard
main sequence T$_{\rm eff}$, \logg,~ and radii 
from Astrophysical Quantities (Cox 1999). 
We repeated the exercise for various
ranges of temperatures and secondary radii.

  Our computed colors show that for late/early type binary configurations,
%namely an early-type main sequence primary paired with a late-type 
%main sequence or giant star, 
we could {\it not} explain the positions of the UVe stars  in a (FUV-NUV), 
$(g-i)_{kis}$ diagram.
% by supposing that the primary and secondary are almost 
%entirely responsible for the visible or UV band colors, respectively. 
In Fig.\,\ref{fngi} we indicate the sequences for G0, F0 and A0 optical
primaries by colored crosses. Our computations show that 
temperature differences between the component binary stars carry the 
(FUV-NUV) and $(g-i)_{kis}$ colors along the main sequence for
single stars - but this is not what is required.
As one moves to models with B type secondaries, the (FUV-NUV) color 
indeed becomes much bluer, but the effect on the $(g-i)$ color is also
to make it too blue for a given shift in (FUV-NUV). As a result the
vector extends approximately along the main sequence locus for single stars.
{\it For no combination of star component temperatures do we find that the 
color vectors move to the lower left (ar even horizontally) and into the UVe 
zone.}
Moreover, if the hot primary is a late-type giant instead of a main sequence 
star, the shifts toward bluer colors quickly become small or insignificant.   
%In sum,  
These and all other trials from our models demonstrated conclusively 
that the colors of binary stars are not moved to the UVe zone.
Accordingly, we must reject a cool-primary, hot-primary binary scenario.
% in which the secondary is a warm or hot star near the main sequence.

\subsubsection{Strong chromospheres?}

\noindent {\it Stellar ages and activity considerations:}

  In active G and K type stars the atmospheric temperature minimum occurs 
at large depths, causing higher temperatures to prevail above this point. 
If a chromospheric temperature rise is strong, it
can influence the formation of strong UV/optical lines. The effects are 
greatest for the principal resonance lines such as
Mg\,II (h/k) and Ca\,II (H/K) resonance lines, and therefore they are the 
first spectral features to be driven into emission. 
Also, enhanced emission of continuum flux first shows its presence
in FUV spectra of young stars.
%According to Findeisen, Hillenbrand, \& Soderblom 2011 (``FHS") and Pagano 
%et al. (2004), although the range of FUV flux can be large at any age the mean
%differences in FUV among like stars in 
%various young clusters (e.g.,~ Blanco 1 at 130\,myr) and Hyades
%(625 Myr) ages can show a spread of as much as 4 magnitudes 
%(it is generally one magnitude or less for the NUV magnitude).

 Several authors have found that the mean FUV emission decreases with
age but levels off by about 1.5\,Gyr (e.g., Pace \& Pasquini 2004, Pace 2013).
The mean NUV decay rate is less certain for older stars. 
Another consideration is that for a given age the enhanced flux from UV 
lines and continuum tends to be weaker as one proceeds along the main sequence 
to M stars (Guinan, Engle, \& DeWarf 2009). 
%In active dMe stars a brightening 
%of line flux in the NUV and FUV bandpasses arises mainly in
%the resonance C\,IV doublet and other prominent emission lines 
%(Robinson et al. 2005, Stelzer 2013). 
%During flares The  (FUV - NUV) color can decrease measurably.  
Although the general dependences of these diagnostics on age have long been 
known broadly, recent work indicates
that critical details of the age-activity-relation are still not understood.
%Indeed our understanding of them can be fairly stated to be in crisis. 
Contrary to earlier discussions,
it is no longer clear that magnetic activity, whether measured by optical 
or X-ray flares, or strong chromospheric emission in the UV, is necessarily 
a reliable indicator of age or vice versa (Soderblom 2010). Indeed, like 
chromospheric strengths can be found in stars that are either very young
or as old as the Hyades.

  Despite these ambiguities ages of stars in the KIC can be estimated
from asteroseismology and from matching the colors and magnitudes to 
up-to-date stellar isochrones. Mathur et al. 
(2013) studied the pulsational mode structure of 22 bright solar-type stars
using different asteroseismological tools and found a uniform distribution of 
ages in the range 3-11 Gyr. 
%However, this was not meant to be a representative
%volume-limited star sample.  
More comprehensively, preliminary results based on
model fittings to isochrones computed from the Dartmouth Stellar Evolution 
Database (Dotter et al. 2008) have been constructed by the
Kepler WG on Stellar Properties (Huber et al. 2014).
This work finds a flat distribution out to 5 Gyr, 
a secondary flat distribution of older stars out to 15 Gyr with an incidence 
of 40\% as great as the incidence of the younger stars. 
A caveat to this description is that below 
0.5 Gyr there is a relative dearth, and then a significant 
increase, of stars with ages of 0.5-1 Gyr. This estimate suggests 
that the incidence of young ($\le$ 0.5 Gyr) stars is 4\%. 
This is about the same as the 3\% 
fraction of a total of 700 solar type stars that exhibit so-called ``very 
active" levels of K line emission (Bastien et al. 2013 and Bastien 2013).
These estimates, albeit preliminary, suggest that the active 
chromospheres hypothesis cannot be dismissed. 
%Moreover, 
%it has not yet been demonstrated that young star chromospheres can 
%regularly produce UV excesses as large as those observed.  
\\

\noindent {\it Light curves and rotational periods:}

Another clue to a G/K star's age is its rotational period or, if no
spot-induced modulation is observable, the presence of a ``flat lined" light 
curve.  To investigate this property we examined Kepler light curves of 55 
``extreme UVe" stars for which (FUV-NUV) and $(g-i)$$_{kis}$ colors are 
$\le$-2.9 and $\ge$0.4 magnitudes, respectively, from those indicated by 
the dot-dashed demarcation line 
in Fig.\,\ref{fngi} and also have Kepler light curves in the MAST archives. 
Twenty nine dwarfs satisfied this criterion. Typical parameters for 
these are T$_{\rm eff}$ $\approx$ 5000\,K and log\,g $\approx$ 4.4.
%\footnote{This ``extreme" sample included 55 UVe dwarfs and 61 (most) 
%of the UVe giants fit this $\Delta$(g-i)$_{kis}$ $>$ criterion. The
%criterion is arbitrarily chosen.}
Contrary to the young active star hypothesis, a substantial fraction 
of these stars exhibit intermediate to long (7--23 day) rotational periods. 
Seven have light curves that can be described as ``flat" or imperceptibly weak 
rotational signals. These are arguably old stars still on the main sequence. 
Light curves of the remaining extreme UVe stars are
chaotic, possibly late-type $\gamma$\,Dor variables or eclipsing binaries. 
About half of these have been studied in detail
(Basri et al. 2010, Prsa 2011, Walkowicz et al. 2011, Howard
et al. 2012), but none has been called out as especially young stars. 
According to our search, most of the UVe stars are not very young. 

 Any results to be drawn from the statistics of binary and 
chromospherically active stars are far from final. Indeed, if we take a 
control sample of UV-normal Kepler-observed dwarfs, we find that their 
light curves are not all flat or dominated by long rotational periods;
many of these stars are pulsators or eclipsing binaries. UV-normal 
G stars with flat curves also exist (e.g., KIC\,11606351), but so
are stars with short rotational periods (e.g., KIC\,11658452; P$_{rot}$
$\approx$ 6 days). These are counterexamples to the proposition that 
moderately rotating G stars must show UV excesses. In all, the light 
curves provide neither refutation nor support for the idea that the 
a young population with active chromospheres is responsible for the
UVe class.
\\

\noindent {\it Enhanced UV flux and X-ray activity:}

\begin{figure}[h!]
\centerline{
\includegraphics[scale=.40,angle=90]{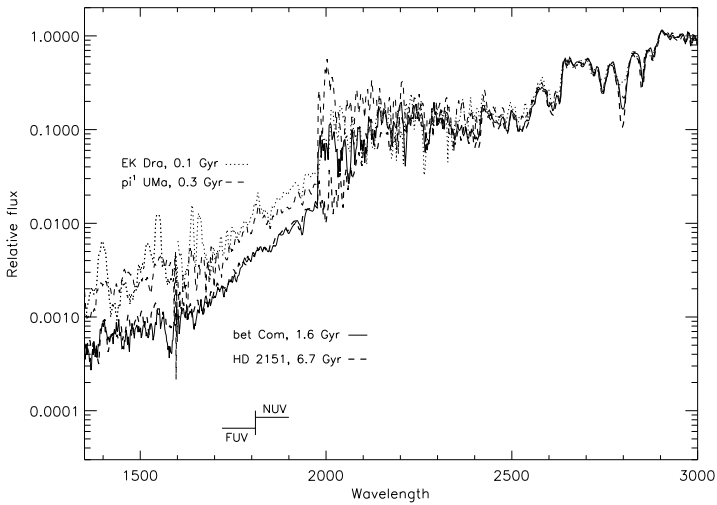}} 
\caption{Smoothed merged IUE spectra of four ``standard" solar-type stars 
of graduated ages. Spectra of the two older stars are represented in bold. 
The GALEX FUV/NUV break is shown.
}
\label{iuesp}
\end{figure}

   Examples of UV color
%significant 
blueing of spectra of young solar-type stars have been
documented (Ribas et al. (2005), Guinan \& Engle (2006), and Linsky et al.
2012).  These examples show that the far-UV continua of young G-M dwarfs can 
be over 3 magnitudes brighter relative to the continua of their oldest stars. 
This is in agreement with the photometric results of Findeisen, 
Hillenbrand, \& Soderblom (2011). In addition,
in Figure\,\ref{iuesp} we exhibit four merged IUE spectra for fluxes for four
solar-type standards (all near G1\,V) and for which ages have been estimated by 
Ribas et al. Here have we scaled the spectra to fluxes of unity at 3000\,\AA.~  
These spectra clearly show a spread of fluxes below $\sim$ 2200\,\AA\,
such that the younger stars have the greatest far-UV fluxes.
% according to Guinan et al. UV flux
Differences can be discerned even between the youngest two stars, 
EK Dra (0.1 Gyr), and $\pi$$^1$\,UMa (0.3 Gyr). 
Our computed (FUV-NUV) colors from their spectra
%, run through the SDSS filter transmission curves and run from 
%youngest to oldest stars, 
are 4.93, 4.70, 6.44, and 6.19. The FUV span is only about 1$\frac{1}{2}$ 
magnitudes.  This is {\it less} than the $\ge$2 magnitudes range in the 
FUV region alone.  Inspection of the spectra suggests that this difference 
is due to a confluence of Fe emission lines at 2000-2100\,\AA\, 
is stronger, e.g., for $\pi$$^1$\,UMa than for the very youngest star, 
EK\,Dra. Similarly, enhanced Fe lines are responsible for a (FUV-NUV) color
inversion of the two older stars, although both FUV and NUV {\it continuum} 
levels maintain a monotonic sequence with age.

%%and according to Ribas et al. increase further
%%for the very youngest and oldest solar stars. 

% FIGGURE 12 ; merged IUE spectra

   To examine the rotational/UV activity arguments further, we have 
cross-correlated the positions of the UVe stars in the Kepler field with 
HEASARC's bright and faint Rosat All Sky Survey catalogs.  Young stars 
with active chromospheres can be expected to exhibit strong soft X-ray flux. 
However, among our UVe sample we find only one solar-type 
star, KIC\,10730406 
%(with KIC T$_{\rm eff}$ = 5070, and log\,g = 3.9), 
that has even a barely significant ($>$3$\sigma$) 
X-ray fluxes detected by the ROSAT surveys. 
(None of our stars were detected in the Chandra or RXTE imaging catalogs 
either, but their coverages of the Kepler fields are very small.)
For a reddening-corrected distance of 700\,pc for KIC\,10730406, the 
implied L$_x$ in the band 0.5-2\,keV is $\sim$3$\times$10$^{30}$\,erg\,s$^{-1}$.
This is consistent with a very young ($\sim$30\,Myr) star with a strong
corona, but not with an older star. Yet many of our stars are
more distant than this one, 
%(which is at an average distance for our UVe sample), 
so it is possible that we might miss several X-ray active stars. 

  It may be a problem to suppose that young stars (e.g., $<$500-600\,Gyr old) 
are even numerous enough to explain the observed population of UVe stars.
%Nor is there reason to believe that some 20\% of our sample are young
% stars (e.g., $<$500-700\,Gyr old) are present at the Galactic latitudes 
The mean age of 3-4 Gyr quoted earlier for solar-type stars implies a more 
or less uniform distribution of ages. This argument carries even more
weight for the UVe stars in our high latitude GALEX-SDSS sample. 
Similarly, the Eclipsing Binary Catalog suffers no known bias toward 
young stars (Prsa, priv. comm.). 
To the contrary, any biases in this catalog should be toward 
underrepresentation of large (evolved) stars in close binaries, not small 
ones on the ZAMS. 
%NOTE: ONLY 1 A. BROWN STAR IS GIVEN AS YOUNG, UV_EXCESS STAR: IT'S KIC7200910,
% AND IT IS JUST BARELY OVER THE DEMARCATION LINE
Finally, those UV excesses that are observed in GK giants
%, which are 10\% of our UVe sample, 
cannot be expected to arise from strong chromospheres. 
All this said, we would not be surprised if strong chromospheres of some 
late-type dwarfs, whether strong from their youth or association in close 
binaries, are the cause behind many or most of the UVe stars. 
%We will attempt to estimate this fraction in the Conclusions.
%Even so, the population
%of young stars with active chromospheres does not alone seem
%high enough to account for the frequency of UVe stars.

\section{Summary and Conclusions}

Spurred by curious features in GALEX-SDSS color diagrams found by
Smith \& Shiao (2011), we have investigated populations of solar-type
stars (F-K dwarfs and some giants) with K$_p$ $\le$
16.0 to demonstrate that at least 17\% of them are observed
with FUV excesses of about 3-4 magnitudes. The existence of these 
excesses is robust, and they are present with nearly the same frequency for
three stellar samples we investigated: (1) the GALEX-SDSS overlap region of the
sky, (2) those overlap regions of the Kepler FOV observed by the GALEX surveys,
(3) the systems listed in the Kepler Eclipsing Binary Catalog. 
%Of these samples, the Kepler FOV sample probably has the strongest 
%magnitude-limited bias. 
Among a smaller number taken from the KOI list, a subsample 
(2 of 12 stars) exhibits UV excesses consistent with these populations.  
GALEX and SDSS spectra of stars show that these excesses diminish from 
the far-UV wavelength region and are still visible in the blue and
even through the visible region.
The distribution of UVe stars in Fig.\,\ref{fngi} suggests that
the $(g-i)$ color is redder than the UV-normal stars that have the same
spectral type and T$_{\rm eff}$. We speculate that the reddening of the
visible band color arises from the protrusion of a strong chromosphere into
the photosphere. The warmer temperature would produce a shift of the primary
continuum contributor from the H$^{-}$ ion to bound-free transitions of 
hydrogen. This in turn shifts the formation of the near infrared continuum 
($i$ filter) deeper into the atmosphere, resulting in relatively more IR flux 
relative to the optical ($g$ filter). 
A similar explanation may hold for the reddening in (r-K) color by UVe stars. 
Thus, the reddening of this color could be consistent with 
the strong chromosphere hypothesis.
 
In all, degenerate star populations (DA and sdOB stars) and especially 
chromospherically active F-K stars could contribute a minor fraction to
the population of UVe stars. We estimate from their occurrence rates a total
of no more than 2\% and 4\%, respectively. 
%from the sizes of their populations
%given in the literature.  In particular, 
From the expectation that the number of young F-G main sequence
stars  is low, we are hard pressed to accept that 
they are numerous to account for the full UVe population by themselves. 
Considering the similar UVe population fractions in the GALEX-SDSS sky 
(Fig.\,\ref{fnteff}), we estimate a floor of perhaps 15-17\% of the 
apparent UVe population. One or more other populations are evidently 
needed to explain their numbers. 
With correction for the FUV incompleteness, this floor falls to 12-14\%.
%Since we rejected the binary scenario in the last section, we are hard-pressed
%at this point to understand how this shortfall can be completely resolved.

 We summarize our evaluation of the contributions of the likeliest
contributors to the UVe population in Table\,1. Our estimates in the table 
for old yet chromospherically active stars and close binaries are taken 
from Pace (2013). Our assessment of the contributions of 
extragalactic and compact objects (WDs, sdBs) come 
%compact objects (WDs, sdBs) comes from 
our assessments in $\S$\ref{discs}. 
We have added halo stars to the latter group as a minor possibility. 
Their expected numbers (Gilmore \& Zeilik 2000) are only about one sixtieth 
of the UVe population. Moreover, their UVe excesses are not as large the as 
3-4 magnitudes observed, and their rotations are slower than the moderate 
ones reported above.  As noted, we estimate 
that roughly 4\% of the F-K stars can be young ($\approx$0.5 Gyr).
Our total of $\ltsim$15\% in the table is already optimistic and yet it 
falls short of our estimate of the observed fraction, 17-22\%. 
However, if we multiply the observed rate by the 
FUV-incompleteness factor 0.82, estimated in $\S$\ref{genproc}, the 
corrected observed UVe rate becomes 14-18\%. This puts us within striking
distance of agreement to the {\rm SUBTOTAL} line in the table.
We represent this correction exercise in Table\,1 by bringing
the predicted component population rates to the observed values, i.e., 
by applying the factor's reciprocal, 1.22, to the {\rm TOTAL} line. 
The agreement is only modestly comforting 
%few of the estimated rates in the table are secure.
%% However, one loose end is the importance of the fact that
%%we are dealing with a magnitude-limited sample. We must boost our total
%%percentage by the estimate of faint UV-normal stars that could not be
%%observed in the FUV but were included in the GALEX sky surveys. This is
%%the correction factor 1.22 noted in $\S$\ref{glxcol} and is included in
%%the penultimate line of our tally in the table. Our final estimate is
%%$\ltsim$15\% $\times$1.22 or $\ltsim$18\%. 
because most of the estimated percentages are upper limits. Hence, we
cannot be assured that we have identified all the components of the UVe
population or their relative proportions.

As this paper was being refereed, A. Brown et al. (2013) reported on a  
%spectroscopic 
study of young G-K\,V bright dwarfs in a 5 deg.$^{2}$ 
Kepler field. They found that spectra of 51 out of 300 stars
in their sample exhibit strong lithium lines and concluded tentatively
that approximate equal numbers of the sample are young single stars and
active binaries. This is consistent with our findings in Table\,1.

% \begin{deluxetable}{lc}  % COMMENTED OUT FOR APJ
%\tablenum{1}    % COMMENTED OUT FOR APJ
%\tablewidth{0pt} % COMMENTED OUT FOR APJ
%\tablecaption{Contributors to the UVe population among F-K stars } % COMMENTED OUT FOR APJ
%\tablehead{   % COMMENTED OUT FOR APJ

%\colhead{Contributing Population} & \colhead{~~~~Estimated Percentage }  \\ % COMMENTED OUT FOR APJ
%  &    \colhead{of Observed Population } \\ % COMMENTED OUT FOR APJ 
%  }   % COMMENTED OUT FOR APJ

%\startdata   % COMMENTED OUT FOR APJ
\begin{table}[h!]   % FOR APJ
\tablenum{1}        % FOR APJ
\begin{center}      % FOR APJ
\center{\caption{Contributors to the UVe population among F-K stars }} % FOR APJ
\begin{tabular}{lc}  % FOR APJ
   \\     % note addition of a blank line here for aesthetics
\tableline\tableline    % FOR APJ
Contributing Population & ~~~~Estimated Percentage \\    % FOR APJ
\tableline    % FOR APJ
Extragalactic &  ~~~~~$<$1     \\
WD, sdB, halo &   ~~~~$\le$2    \\ 
Old active Stars &  ~~~~$\le$5     \\
Active binaries & ~~~~$\le$4    \\
Young stars & ~~~$\approx$4   \\
%AF + GK\,III  & ~~~$\le$5   \\
 &     \\
%SUBTOTAL: & ~~~~$\le$15\%    \\
SUBTOTAL: & ~~~~$\le$16\%    \\
%      &                   \\
FUV incompleteness corr.: & ~~~1.22$\times$ \\
%TOTAL:  & ~~~~$\le$18\%    \\
TOTAL:  & ~~~~$\le$20\%    \\
\tableline % NEW, for APJ
\end{tabular} % NEW, for APJ
\end{center} % NEW, for APJ
\end{table} % NEW, for APJ
%\enddata \end{deluxetable} % COMMENTED OUT FOR APJ

The existence of the UVe population influences the 
mean UV colors of another stellar group. 
Our UVe group consists mainly of late F to G and K stars. Yet
the A/early-F stars, especially giants. are not among them. 
The AF stars constitute a distinct ``UV blue" population because of 
their higher effective temperatures.
%when one examines the (FUV-NUV) distribution of all the stars. 
In contrast, the UVe frequency is higher among GK and early M-type main 
sequence stars. This strongly pulls their mean (FUV-NUV) colors
to the blue (Fig.\,\ref{fndist}). 
These late-type stars comprise the UV-blue stars in Fig.\,\ref{ging1}, 
and thus they appear to surround a ``UV-red clump."
%This effect is much smaller on the giants,
%indicating that the UV flux contamination among comparable UVe systems is
%much weaker. This argues against the strong chromospheres hypothesis for
%those giants tha have UV excesses since 
%giants are chromospherically quiet.
%, and the fraction of higher luminous young (T Tauri) stars is effectively 
%zero in this population. 
%On the other side of the ledger, a strong countervailing argument against the 
%binary hypothesis is the obvious fact that the evolution and brightening of 
%stars across the lower main sequence band starts with a slow increasing of 
%the effective temperature followed by its more rapid decrease as the 
%star transits toward the red giant branch.
% particularly compelling evidence in our view that this component is a
% hotter star on the main sequence but which is relatively subluminous 
% owing to the brightening of the primary as it turns off the main 
% sequence.
%The stars exhibiting blue UV colors ($FUV-NUV$ $<$ 2.5) according to
%the flux distributions expected for their T$_{\rm eff}$s as single stars.
%The effect of the UV excess on the binaries with cooler secondaries pulls 
%the mean colors of these spectral types to the blue, responsible for the 
%{\it reversal} of the UV color back to the blue as one moves from hot to 
%warm to finally cool stars. 
To complete this discussion, notice that the {\it blue} peak of the 
histogram in Fig.\,\ref{fndist} is significantly augmented by the addition of 
{\it } faint, point-like extragalactic objects (Bianchi et al. 2007). This can 
be seen in the 17$<$ K$_p$ $<$ 19 mag. distribution broken out in this figure.

We look forward to detailed studies of the EBC systems and 
particularly KOI stars with UVe colors. These can determine if
near-UV fluxes from the secondaries influence the colors
of the primaries as they eclipse them. 
%parameters of the primary stars and, from transit depths, any planets 
%they may host. 
We also anticipate that radial velocity studies of many of these
stars will refine the frequency estimates of active binaries.
A study of the ages of these objects via 
asteroseismology will also address whether young active chromospheres 
can be expected to be prevalent after all. 
In that event one can hope to amass a sample of young planet-hosting stars
to compare with older stars to find out how planet-bearing incidence changes
with time and also to map the evolution of planetary system properties.
%Ironically, and for quite different reasons,
%according to both the binary or active 
%chromosphere scenarios the issue is that we do not quite understand
%yet important properties of very young, late-type stars.
%% (whether more massive binaries or solar-type single stars).

\acknowledgments
We thank Dr. R. Ostensen for his sending us his catalog of white dwarfs. 
We appreciate also Dr. D. Huber's permission to quote details of work 
done by the Kepler Working Group on Stellar Properties 
%in advance of its publication 
and Dr. F. Bastien's notice of his K-line work. 
This paper has profitted from valuable comments by the referee.

\end{document}